\documentclass[aps,floats]{revtex4}
\usepackage{amsmath,amssymb}
\usepackage{graphicx,epsfig}

\begin{document}
\bibliographystyle {plain}

\def\oppropto{\mathop{\propto}} 
\def\opsimeq{\mathop{\simeq}}
\def\opoverderline{\mathop{\overline}}
\def\operarrow{\mathop{\longrightarrow}}
\def\opsim{\mathop{\sim}}

\def\fig#1#2{\includegraphics[height=#1]{#2}}
\def\figx#1#2{\includegraphics[width=#1]{#2}}


\title{  Typical versus averaged overlap distribution in Spin-Glasses : \\
 Evidence for the droplet scaling theory
 } 


 \author{ C\'ecile Monthus and Thomas Garel }
  \affiliation{ Institut de Physique Th\'{e}orique, CNRS and CEA Saclay,
 91191 Gif-sur-Yvette, France}

\begin{abstract}
We consider the statistical properties over disordered samples $(\cal J)$ of the overlap distribution $P_{\cal J}(q)$ which plays the role of an order parameter in spin-glasses. We show that near zero temperature (i) the {\it typical } overlap distribution is exponentially small in the central region of $-1<q<1$: $ P^{typ}(q) = e^{\overline { \ln P_{\cal J}(q) }} \sim e^{ - \beta N^{\theta} \phi(q)} $, where $\theta$ is the droplet exponent defined here with respect to  the total number $N$ of spins (in order to consider also fully connected models where the notion of length does not exist); (ii) the  rescaled variable $v = - (\ln P_{\cal J}(q))/N^{\theta}$ remains an $O(1)$ random positive variable describing sample-to sample fluctuations; (iii) the averaged distribution $\overline{P_{\cal J}(q) } $ is non-typical and dominated by rare anomalous samples. Similar statements hold for the cumulative overlap distribution $I_{\cal J}(q_0) \equiv \int_{0}^{q_0} dq P_{\cal J}(q) $. These results are derived explicitly for the spherical mean-field model with $\theta=1/3$, $\phi(q)=1-q^2 $, and the random variable $v$ corresponds to the rescaled difference between the two largest eigenvalues of GOE random matrices. Then we compare numerically the typical and averaged overlap distributions for the long-ranged one-dimensional Ising spin-glass with random couplings decaying as $J(r) \propto r^{-\sigma}$ for various values of the exponent $\sigma$, corresponding to various droplet exponents $\theta(\sigma)$, and for the mean-field SK-model (corresponding formally to the $\sigma=0$ limit of the previous model). Our conclusion is that future studies on spin-glasses should measure the {\it typical} values of the overlap distribution $P^{typ}(q)$ or of the cumulative overlap distribution $I^{typ}(q_0)= e^{\overline { \ln I_{\cal J}(q_0) }}$ to obtain clearer conclusions on the nature of the spin-glass phase.

\end{abstract}

\maketitle

\section{ Introduction } 

In the statistical physics of quenched disordered systems, 
where each disordered sample $(\cal J)$ is characterized by its partition function
\begin{eqnarray}
 Z_{\cal J} && = \sum_{\cal C} e^{- \beta E_{\cal J}( \cal C)} = e^{- \beta F_{\cal J}(\beta)}
\label{Z}
\end{eqnarray}
it has been realized from the very beginning \cite{brout}
that the quenched free-energy 
\begin{eqnarray}
\overline{ \ln Z_{\cal J} } && = - \beta \overline{ F_{\cal J}} 
\label{Ztyp}
\end{eqnarray}
is typical, i.e. is representative of the physics in almost all samples $(\cal J)$,
whereas the averaged partition function $\overline{  Z_{\cal J} }$ 
can be non-typical, especially at low temperature,
because it can be dominated by very rare disordered samples $(\cal J)$.
Correlation functions are, from this point of view, very similar to
partition functions : the averaged correlation can be very different from the typical
correlation. It is very clear in one dimensional spin
systems \cite{Der_Hil,luck}, where correlation functions can be
written as product of random numbers, but it is also true 
for higher dimensional models  \cite{mccoywu,danielrtfic,danielreview,Lud,ber_cha}.
More generally, for each observable, it is very important to be aware of the
possible differences between typical and averaged values, and to have a clear idea of the
distribution over samples. 

In the field of classical spin-glasses (see for instance \cite{binder_young,young,newman_stein_book}), 
there has been an ongoing debate on the nature of the spin-glass phase between 
the droplet scaling theory \cite{mcmillan,bray_moore,fisher_huse}, which is based on real space renormalization ideas (explicit real-space renormalization for spin-glasses have been studied in detail within the Migdal-Kadanoff approximation \cite{hierarchicalspinglass}),
  and the alternative Replica-Symmetry-Breaking scenario \cite{replica}
based on the mean-field fully connected Sherrington-Kirkpatrick model \cite{SKmodel}.
The questions under debate include the presence of the number of ground states (two or many)
\cite{FH_purestates,newman_stein,pala_pure}, the properties of the overlap 
\cite{overlap_drossel,overlap_martin,overlap_pal,overlap_young,overlap_domany,overlap_aspelmeier,overlap_banos,overlap_fernandez,overlap_perco}, 
the statistics of excitations 
\cite{excita_3d,excita_3dbis}, the structure of state space \cite{domany},
the absence or presence of an Almeida-Thouless line in the presence of an magnetic field
\cite{almeida3d_martin,almeida3d_young,almeida3d_jorg,almeida_MB,almeida_Katz,almeida_temesvari}, etc ...
In particular, one of the standard observable to discriminate between the droplet and the replica theories
has been the {\it averaged } overlap distribution $\overline{P_{\cal J}(q)} $.
In the present paper, we show that this {\it averaged } overlap distribution $\overline{P_{\cal J}(q)} $ 
is actually non-typical and is governed by rare disordered samples, whereas the {\it typical } overlap distribution 
\begin{eqnarray}
P^{typ}(q) \equiv e^{\overline {\ln P_{\cal J}(q)}}
\label{overtyp}
\end{eqnarray}
 is in full agreement with the droplet scaling theory.
Our conclusion is that it does not seem a good idea to use a non-typical observable such as the {\it averaged } overlap distribution $\overline{P_{\cal J}(q)} $ to elucidate the physics of spin-glasses, 
and that future studies
should focus on  the  {\it typical } overlap distribution to obtain clear conclusions.
Note that two recent studies have also proposed to study other statistical properties of the overlap
distribution $P_{\cal J}(q) $ than the averaged value, namely the statistics of peaks \cite{yucesoy}
or the median over samples of cumulative overlap distribution \cite{middleton}.
We hope that the numerical measure of the {\it typical } overlap distribution, which is a much simpler 
observable, will give even clearer evidence for the droplet scaling theory.

The paper is organized as follows.
In section \ref{sec_overlap}, we discuss the general properties of the overlap distribution.
In section \ref{sec_sph}, we derive explicit results for the spherical mean-field model.
In section \ref{sec_LRSG}, we present numerical results for the one-dimensional  long-ranged spin-glass
with random couplings decaying as $J(r) \propto r^{-\sigma}$ for various values of the exponent $\sigma$.
In section \ref{sec_SK}, we show numerical results for the mean-field SK-model
(corresponding formally to the $\sigma=0$ limit of the previous model).
Our conclusions are summarized in section \ref{sec_conclusion}.
The Appendix \ref{app_theta} contains a brief reminder on the physical meanings
 of the droplet exponent $\theta$, whereas Appendix \ref{app_rep} briefly recalls
the replica prediction for the distribution of the cumulative overlap distribution.

\section{ Overlap distribution  in a given disordered sample} 

\label{sec_overlap}

\subsection{ Notations } 

Let us consider a general spin-glass model containing $N$ spins $S_i=\pm 1$ and random couplings 
${\cal J} \equiv \{J_{ij} \}$
\begin{eqnarray}
 H_{\cal J} && = - \sum J_{ij} S_i S_j
\label{HSG}
\end{eqnarray}
The partition function associated to the disordered sample $ {\cal J} \equiv \{J_{ij} \}$ reads
\begin{eqnarray}
 Z_{\cal J}^{single}(\beta) && = \sum_{\{S_i=\pm 1\}} e^{\beta \sum J_{ij} S_i S_j } 
\label{Z1SG}
\end{eqnarray}
We use here the notation 'single' to stress that this partition function contains
a single 'copy' of spins, in contrast to partition functions concerning 'two copies' of spins
that we will introduce below.
To characterize the spin-glass 'order', one introduces the overlap
\begin{eqnarray}
Q=  \sum_{i=1}^N S_i^{(1)} S_i^{(2)} 
\label{defq}
\end{eqnarray}
between two independent copies  of spins $(S_i^{(1)}=\pm 1,S_i^{(2)}=\pm 1)$
in the same disordered sample ${\cal J} \equiv \{J_{ij} \} $. 
The parameter $Q$ of Eq. \ref{defq} can take the $(N+1)$ discrete values $-N,-N+2,...,N-2,N$,
so that for large system it is convenient to consider the rescaled overlap 
\begin{eqnarray}
q \equiv \frac{Q}{N}
\label{defqrescal}
\end{eqnarray}
which remains in the interval $-1 \leq q \leq 1$.

\subsection{ Overlap distribution as a ratio of partition functions } 

The probability distribution of the overlap introduced in Eq. \ref{defq}
can be written as the ratio of two partition functions
concerning the two copies
\begin{eqnarray}
 {\cal P}_{\cal J}(Q)= \frac{ {\cal Z}_{\cal J}(\beta;Q)}{ {\cal Z}_{\cal J}(\beta)}
\label{defpq}
\end{eqnarray}
The numerator of Eq. \ref{defpq} represents the partition function of two copies in the same disorder
constrained to a given overlap $Q$ (Eq \ref{defq})
\begin{eqnarray}
 {\cal Z}_{\cal J}(\beta;Q) \equiv \sum_{\{S_i^{(1)}=\pm 1\}} \sum_{\{S_i^{(2)}=\pm 1\}}
e^{\beta \sum J_{ij} (S_i^{(1)} S_j^{(1)}+S_i^{(2)} S_j^{(2)}) } \delta_{Q,  \sum_{i=1}^N S_i^{(1)} S_i^{(2)}}
\label{numepq}
\end{eqnarray}
The denominator is the full partition function of the two copies in the same disorder,
with no constraint on the overlap,
so that it factorizes into the product of two partition functions concerning a single copy
(Eq \ref{Z1SG})
\begin{eqnarray}
 {\cal Z}_{\cal J}(\beta) \equiv \sum_{\{S_i^{(1)}=\pm 1\}} \sum_{\{S_i^{(2)}=\pm 1\}}
e^{\beta \sum J_{ij} (S_i^{(1)} S_j^{(1)}+S_i^{(2)} S_j^{(2)}) } = (Z_{\cal J}^{single}(\beta) )^2 
\label{denopq}
\end{eqnarray}

The fact that the overlap distribution ${\cal P}_{\cal J}(Q)$ is a ratio of two partition functions (Eq. \ref{defpq}) yields that its logarithm $\ln {\cal P}_{\cal J}(Q)$ corresponds to a difference
of two free-energies
\begin{eqnarray}
 \ln {\cal P}_{\cal J}(Q) = \ln {\cal Z}_{\cal J}(\beta;Q) -   \ln {\cal Z}_{\cal J}(\beta) 
\label{overtypdiffFinter}
\end{eqnarray}
Since averaged free-energies are known to be typical (see the Introduction around Eq. 2),
the typical overlap distribution defined as
\begin{eqnarray}
\ln P^{typ}(Q) \equiv \overline{ \ln {\cal P}_{\cal J}(Q)}= \overline{ \ln {\cal Z}_{\cal J}(\beta;Q) }
-  \overline{ \ln {\cal Z}_{\cal J}(\beta) }
\label{overtypdiffF}
\end{eqnarray}
will be representative of most samples, whereas the averaged value $P^{av}(Q) $ 
obtained by averaging directly 
the ratio of partition functions of Eq. \ref{defpq}
\begin{eqnarray}
 P^{av}(Q) \equiv \overline{ {\cal P}_{\cal J}(Q) }
=  \overline{ \left( \frac{ {\cal Z}_{\cal J}(\beta;Q)}{ {\cal Z}_{\cal J}(\beta)} \right) }
\label{defpqav}
\end{eqnarray}
can be dominated by non-typical disordered samples, especially at very low temperature
as we now discuss.

\subsection{ Behavior near zero temperature }

Exactly at zero temperature, the single-copy partition function of Eq. \ref{Z1SG}
will be dominated by the ground-state energy $E_{\cal J}^{(GS)}$ corresponding to the two ground states
related by a global flip of all the spins ($\{S_i^{(GS)}\}$ and  $\{-S_i^{(GS)}\}$)
\begin{eqnarray}
 Z_{\cal J}^{single}(\beta) && \opsimeq_{T \to 0} 2 e^{- \beta E_{\cal J}^{(GS)}}
\label{Z1SGtzero}
\end{eqnarray}
The two-copies partition function of Eq. \ref{numepq} will also be dominated 
by the cases where each of the two copies is in either of the two ground-states, so that it reads
\begin{eqnarray}
 {\cal Z}_{\cal J}(\beta;Q) \opsimeq_{T \to 0} e^{- 2 \beta E_{\cal J}^{(GS)} } \left( 2 \delta_{Q, N} + 2 \delta_{Q, -N} \right)
\label{numepqtzero}
\end{eqnarray}
The overlap distribution of Eq. \ref{defpq} has thus the following expected
 zero-temperature limit in each sample
\begin{eqnarray}
 {\cal P}_{\cal J}^{T=0}(Q)= \frac{1}{2} \left(  \delta_{Q, N} +  \delta_{Q, -N} \right)
\label{pqtzero}
\end{eqnarray}

To obtain the dominant contribution near zero temperature at a given overlap value $Q \ne \pm N$,
we may consider that one of out the two copies (say $S_i^{(1)}$) is in one of the ground-states (say $(S_i^{(GS)})$)
in Eq. \ref{numepq} : then to obtain a given overlap $Q$, the second copy $S_i^{(2)}$ must have
\begin{eqnarray}
n \equiv \frac{N-Q}{2}
\label{defp}
\end{eqnarray}
 spins different from the first copy ($S_i^{(2)}=-S_i^{(1)}$) and $N-n=(N+Q)/2$ spins identical to the first copy ($S_i^{(2)}=S_i^{(1)}$)
\begin{eqnarray}
 {\cal Z}_{\cal J}(\beta;Q = N-2n ) && \simeq 4  \sum_{1 \leq i_1 <i_2 < .. < i_n \leq N} e^{- \beta E_{\cal J}(i_1,..,i_n) }
\nonumber \\
E_{\cal J}(i_1,..,i_n) && \equiv - \sum_{ij} J_{ij} S_i^{(2)}  S_j^{(2)} \left( \prod_{k=1}^n \delta_{S_{i_k}^{(2)},-S_{i_k}^{GS}} \right)
\prod_{i \ne (i_1,..,i_n)} \delta_{S_{i_k}^{(2)},S_{i_k}^{GS}}
\label{numepqleading}
\end{eqnarray}
The ratio of Eq. \ref{defpq} for $Q \ne \pm N$ will thus have for leading contribution
\begin{eqnarray}
 {\cal P}_{\cal J}(Q= N-2n) && \simeq  \sum_{1 \leq i_1 <i_2 < .. < i_n \leq N} e^{- \beta (E_{\cal J}(i_1,..,i_n)-E_{\cal J}^{GS}) }
\label{pqtnearzero}
\end{eqnarray}
which represents the partition function of excitations of a given size $n$.
 Near zero temperature, one further expects that in each given sample, the overlap distribution
will be dominated by the biggest of these contributions
\begin{eqnarray}
 {\cal P}_{\cal J}(Q= N-2n) && \simeq    e^{- \beta E_{\cal J}^{min}(n) }
\label{pqtnearzeromin}
\end{eqnarray}
where 
\begin{eqnarray}
E_{\cal J}^{min}(n) \equiv  \min_{1 \leq i_1 <i_2 < .. < i_n \leq N} \left(E_{\cal J}(i_1,..,i_n)-E_{\cal J}^{GS}\right) 
\label{eminp}
\end{eqnarray}
represents the minimal energy cost $(E_{\cal J}(i_1,..,i_n)-E_{GS})$ among all 
excitations involving the flipping of exactly $n=\frac{N-Q}{2}$ spins with respect to the ground state.

So we expect that the typical overlap has the following leading behavior near zero temperature
\begin{eqnarray}
\ln {\cal P}^{typ}(Q) \equiv \overline{\ln {\cal P}_{\cal J}(Q)} && \simeq   - \beta \overline{ E_{\cal J}^{min}(n=\frac{N-Q}{2}) }
\label{pqtnearzeromintyp}
\end{eqnarray}

\subsection{ Relation with the droplet scaling theory  } 

The probability distribution $P_{\cal J}(q)$ of the rescaled variable $q=Q/N$ of Eq. \ref{defqrescal}
reads near zero temperature (Eq. \ref{pqtnearzeromin}) 
\begin{eqnarray}
P_{\cal J}(q) = N {\cal P}_{\cal J}(Q) \simeq  e^{- \beta E_{\cal J}^{min}\left(n=N \frac{(1-q)}{2}\right) }
\label{defpqrescal}
\end{eqnarray}
In the central region $-1<q<1$, the number $n=N \frac{(1-q)}{2}$ of spins is extensive in the total number $N$ of spins of the disordered sample. According to the droplet scaling theory \cite{mcmillan,bray_moore,fisher_huse}, the droplet exponent $\theta$ describes the scaling of the energy 'optimized excitations'
with respect to their size (see Appendix \ref{app_theta}), so that we expect the scaling
\begin{eqnarray}
\ln P_{\cal J}(q) \simeq - \beta E_{\cal J}^{min}\left(n=N \frac{(1-q)}{2}\right)  \simeq - \beta N^{\theta} v
\label{defpqrescaltheta}
\end{eqnarray}
where $v$ is a positive random variable of order $O(1)$.
In particular, the corresponding typical value is exponentially small
\begin{eqnarray}
\ln {\cal P}^{typ}(q) \equiv \overline{\ln {\cal P}_{\cal J}(q)}  \simeq - \beta N^{\theta} 
\label{defpqrescalthetatyptyp}
\end{eqnarray}
whereas the averaged value will be governed by the rare samples having an anomalous
small variable $v \leq T/N^{\theta}$.
This analysis leads to a power-law decay with respect to the size $N$
\begin{eqnarray}
P^{av}(q) \equiv \overline{ P_{\cal J}(q) } \propto N^{-x}
\label{probenav}
\end{eqnarray}
where the exponent $x$ depends on the behavior of the probability distribution of the variable $v$ near the origin $P(v \to 0)$,
as well as on possible prefactors in front of the exponential factor of Eq. \ref{defpqrescal}.
For short-ranged spin-glass models, the standard droplet scaling theory
\cite{mcmillan,bray_moore,fisher_huse} predicts a finite weight at the origin $P(v=0)>0$
for the variable $v$, and no size-prefactors, 
so that the exponent $x$ takes the simple value given by the droplet exponent 
\begin{eqnarray}
x_{simple}=\theta
\label{xsimple}
\end{eqnarray}
 However it is clear that these are two additional properties
with respect to the analysis of the typical behavior. For instance, in the quantum random transverse-field Ising chain \cite{danielrtfic}, equivalent to the two dimensional classical McCoy-Wu model \cite{mccoywu}, the typical correlation function decays as $C_{typ}(r) = e^{\overline{ \ln C(r)}} \sim e^{- r^{\theta} u}$ with the simple droplet exponent $\theta=1/2$, whereas the {\it averaged} correlation decays
as the power-law $\overline {C(r)} \propto r^{-x}$ with the non-trivial exponent $x = (3-\sqrt{5})/2$
\cite{danielrtfic}.
In summary, we feel that the exponential typical decay of Eq. \ref{defpqrescalthetatyptyp}
is a very robust conclusion of the droplet scaling theory, whereas the power-law decay with 
$x_{simple}=\theta $ of the averaged value is based on further hypothesis that are less general
(see for instance the section \ref{sec_sph} concerning the spherical model where the variable $v$
does not have a finite weight near the origin (Eq. \ref{vsph})).

\subsection{ Cumulative overlap distribution in each sample } 

It is convenient to consider also the cumulative overlap distribution
\begin{eqnarray}
I_{\cal J}(q_0) && \equiv \int_{0}^{q_0} dq P_{\cal J}(q) 
= \sum_{Q=0}^{N q_0} {\cal P}_{\cal J}(Q)
\label{defcumulative}
\end{eqnarray}
Near zero temperature, the leading contribution of Eq. \ref{pqtnearzero} yields
\begin{eqnarray}
I_{\cal J}(q_0)  \simeq \sum_{n=\frac{N (1-q_0)}{2}  }^{\frac{N}{2}} \ \ \ \sum_{1 \leq i_1 <i_2 < .. < i_n \leq N} e^{- \beta (E_{\cal J}(i_1,..,i_n)-E_{\cal J}^{GS}) }
\label{iqtnearzero}
\end{eqnarray}
which represents the partition function over excitations containing $n$ flipped spins
with respect to the ground state, where $n$ is in the interval $ \frac{N (1-q_0)}{2} \leq n \leq \frac{N}{2}$. The important point is that the minimal value $\frac{N (1-q_0)}{2} $ is also system-size.
So from the point of view of the droplet scaling theory, the minimal energy cost
of these system-size excitations in each sample will lead to the same scaling as Eq. \ref{defpqrescaltheta}
\begin{eqnarray}
\ln I_{\cal J}(q_0) \simeq - \beta E_{\cal J}^{min}\left(N \frac{(1-q_0)}{2} \leq n \leq \frac{N}{2} \right)  \simeq - \beta N^{\theta} v
\label{cumulativerescaltheta}
\end{eqnarray}
where $v$ is a positive random variable of order $O(1)$.
As a consequence, the typical value $ I^{typ}(q_0) $ will be exponentially small
\begin{eqnarray}
\ln I^{typ}(q_0) \equiv \overline{\ln I_{\cal J}(q_0)} \simeq  - \beta N^{\theta} 
\label{cumulativetyptheta}
\end{eqnarray}
On the contrary, within the replica theory \cite{replica}, the typical
value remains finite for $N =+\infty$ (see Eq. \ref{repityp} of Appendix B),
i.e. roughly speaking, this corresponds to a vanishing droplet exponent $\theta=0$.

Again, Eq. \ref{cumulativerescaltheta} yields that the averaged value  $I^{av}(q_0)$
of the cumulative distribution
will be governed by the rare samples having an anomalous small variable $v$
and will decay as a power-law as Eq. \ref{probenav}.

\section{ Fully connected Spherical Spin-Glass model } 

\label{sec_sph}

In this section, we consider the fully connected Spherical Spin-Glass model introduced in \cite{sphericalSG}
defined by the Hamiltonian
\begin{eqnarray}
H_{\cal J}= - \sum_{ 1 \leq i < j\leq N } J_{i,j} S_i S_j = - \frac{1}{2} \sum_{ i \ne j} J_{i,j} S_i S_j
\label{Hsph}
\end{eqnarray}
where the random couplings $J_{i,j}=J_{j,i}$ are drawn with the Gaussian distribution
\begin{eqnarray}
P(J_{ij}) = \sqrt{ \frac{N }{2 \pi  } }  e^{ - \frac{N J_{ij}^2 }{ 2  }}
\label{gausssph}
\end{eqnarray}
and where the spins are not Ising variables $S_i=\pm 1$ but are instead continuous variables $S_i \in ]-\infty,+\infty[$ submitted to the global constraint
\begin{eqnarray}
\sum_{i=1}^N S_i^2=N
\label{contraintesph}
\end{eqnarray}
so that the partition function for a given sample reads
\begin{eqnarray}
Z_{\cal J}^{single}(\beta)  = \left( \prod_{i=1}^N \int_{-\infty}^{+\infty} dS_i \right) e^{
\frac{\beta}{2} \sum_{ i\ne j} J_{i,j} S_i S_j}  \delta( N - \sum_{i=1}^N S_i^2 )
\label{zsph}
\end{eqnarray}

\subsection{  Ground state energy in each sample } 

The random couplings $J_{ij}$ form a random Gaussian symmetric matrix $\tilde J$ of size $N$.
Let us introduce its $N$ eigenvalues in the order
\begin{eqnarray}
\lambda_1>\lambda_2>..>\lambda_N
\label{diagolambda}
\end{eqnarray}
and the corresponding basis of eigenvectors $\vert \lambda_p >$ to have the spectral decomposition
\begin{eqnarray}
{\tilde J} = \sum_{p=1}^N \lambda_p \vert \lambda_p >< \lambda_p \vert
\label{diagoj}
\end{eqnarray}
Writing the spin vector in this new basis
\begin{eqnarray}
\vert S > = \sum_{i=1}^N S_i \vert i > = \sum_{p=1}^N S_{\lambda_p} \vert \lambda_p >
\label{sdiagoj}
\end{eqnarray}
the partition function of Eq. \ref{zsph} becomes
\begin{eqnarray}
Z_{\cal J}^{single}(\beta)  = \left( \prod_{p=1}^N \int_{-\infty}^{+\infty} dS_{\lambda_p} \right) 
e^{ \frac{\beta}{2} \sum_{p=1}^{N} \lambda_p S_{\lambda_p}^2 }
  \delta( N - \sum_{p=1}^N S_{\lambda_p}^2 )
\label{zsphdiag}
\end{eqnarray}
The ground-state is now obvious : to maximize the argument of the exponential,
one needs to put the maximal possible weight in the first possible maximal eigenvalue $\lambda_1$
(Eq. \ref{diagolambda})
and zero weight in all other eigenvalues $\lambda_p$ with $p=2,3,..,N$
\begin{eqnarray}
S_{\lambda_p \ne \lambda_1}^{GS} && = 0
\nonumber \\
S_{\lambda_1}^{GS} && = \pm \sqrt{N- \sum_{p=2}^N (S_{\lambda_p}^{GS})^2} = \pm \sqrt{N}
\label{eigengs}
\end{eqnarray}
So Eq \ref{zsphdiag} has for leading exponential term
\begin{eqnarray}
Z_{\cal J}^{single}(\beta)  \propto 
e^{ \frac{\beta}{2}  \lambda_1 (S_{\lambda_p}^{GS})^2 } =e^{ \beta \frac{N  \lambda_1 }{2}   }
\label{zsphzeroT}
\end{eqnarray}
and the ground-state energy is simply determined by the first eigenvalue $\lambda_1$
\begin{eqnarray}
E^{GS}_{\cal J}(N) \simeq \frac{ \ln Z_{\cal J}^{single}(\beta) }{ (- \beta) } = - N \frac{ \lambda_1 }{2}
\label{egssph}
\end{eqnarray}
The statistics of the largest eigenvalue $\lambda_1$ of random Gaussian symmetric matrices
(Gaussian-Orthogonal-Ensemble) is known to be given by
\begin{eqnarray}
 \lambda_1 = 2  \left( 1+ \frac{u}{2 N^{2/3}} \right)
\label{lambdamaxTW}
\end{eqnarray}
where the value $2 $ corresponds to the boundary of the semi-circle law that emerges in the thermodynamic limit
$N \to +\infty$, and where $u$ is a random variable of order $O(1)$
distributed with the Tracy-Widom distribution \cite{tracyWidom}.
The ground-state energy thus reads
\begin{eqnarray}
E^{GS}_{\cal J}(N)= - N \frac{ \lambda_1 }{2} = -  N - N^{1/3} \frac{ u}{2 }
\label{egssphfinal}
\end{eqnarray}
In summary, the extensive term is non-random, and the next subleading term
is of order $N^{1/3}$ and random, distributed with
 the Tracy-Widom distribution, as already mentioned in \cite{andreanov}.
Within the general analysis of the statistics of the ground state energy recalled in the Appendix
 (Eqs \ref{e0av} and \ref{e0fluct}), this means that the spherical model has for droplet exponent and for fluctuation exponent the same simple value
\begin{eqnarray}
\theta^{sph} && =\frac{1}{3}
\nonumber \\
\mu^{sph} && =\frac{1}{3}
\label{thetasph}
\end{eqnarray}

\subsection{ Overlap distribution in each sample }

To analyze the overlap distribution in a given sample,
we analyze similarly the two-copies partition function (Eq. \ref{numepq})
\begin{eqnarray}
 {\cal Z}_{\cal J}(\beta ;  Q) = 
&& \left( \prod_{i=1}^N \int_{-\infty}^{+\infty} dS^{(1)}_i \int_{-\infty}^{+\infty} dS^{(2)}_i\right)
  e^{
\frac{\beta}{2} \sum_{ i\ne j} J_{i,j} (S^{(1)}_i S^{(1)}_j  +S^{(2)}_i S^{(2)}_j ) } 
  \nonumber \\ &&  \delta( N - \sum_{i=1}^N (S_i^{(1)})^2  )
 \delta( N - \sum_{i=1}^N (S_i^{(2)})^2  )
 \delta \left(Q -  \sum_{i=1}^N S_i^{(1)} S_i^{(2)} \right)
\label{numepqsphd}
\end{eqnarray}
Using the basis of eigenvectors of the matrix ${\tilde J}$ of the couplings (Eq. \ref{diagoj}),
Eq. \ref{numepqsphd} becomes
\begin{eqnarray}
 {\cal Z}_{\cal J}(\beta ;  Q) && = \left( \prod_{p=1}^N \int_{-\infty}^{+\infty} dS_{\lambda_p}^{(1)} 
\int_{-\infty}^{+\infty} dS_{\lambda_p}^{(2)} \right) 
e^{ \frac{\beta}{2}
\sum_{p=1}^{N} \lambda_p \left( (S_{\lambda_p}^{(1)})^2 +(S_{\lambda_p}^{(2)})^2 \right) }
  \nonumber \\ &&  \delta( N - \sum_{p=1}^N (S_{\lambda_p}^{(1)})^2  )
 \delta( N - \sum_{p=1}^N (S_{\lambda_p}^{(2)})^2  )
 \delta \left(Q -  \sum_{p=1}^N S_{\lambda_p}^{(1)} S_{\lambda_p}^{(2)} \right)
\label{numepqsphdiag}
\end{eqnarray}
To obtain the leading behavior near zero temperature, we may consider that one of the copy
(say $S^{(1)}$) is in one of the two ground-states (Eq \ref{eigengs})
\begin{eqnarray}
S_{\lambda_p \ne \lambda_1}^{(1)}&& =S_{\lambda_p \ne \lambda_1}^{GS}  = 0
\nonumber \\
S_{\lambda_1}^{(1)} && =S_{\lambda_1}^{GS}  =  \sqrt{N}
\label{eigengs1}
\end{eqnarray}
Then the component $S_{\lambda_1}^{(2)}$ of the second copy on the first eigenvector
is completely fixed by the overlap $Q$
\begin{eqnarray}
 S_{\lambda_1}^{(2)} =\frac{Q}{ \sqrt{N}}
\label{composante2gs}
\end{eqnarray}
The best that we can do for the second copy is thus to put all the remaining weight on the second eigenvalue, and zero weight on higher eigenvalues $p=3,4,..,N$
\begin{eqnarray}
S_{\lambda_p < \lambda_2}^{(2)}&& = 0
\nonumber \\
 S_{\lambda_2}^{(2)}&&  =\sqrt{ N-( S_{\lambda_1}^{(2)})^2 } = \sqrt{ N- \frac{Q^2}{ N}}
\label{composantelambda2}
\end{eqnarray}
Then the leading exponential term of Eq. \ref{numepqsphdiag} reads
\begin{eqnarray}
 {\cal Z}_{\cal J}(\beta ;  Q) && \propto   
e^{ \frac{\beta}{2} \left[
 \lambda_1 \left( (S_{\lambda_1}^{(1)})^2 +(S_{\lambda_1}^{(2)})^2 \right)
+ \lambda_2 (S_{\lambda_2}^{(2)})^2 \right] }
   \nonumber \\ && 
=  e^{ \frac{\beta}{2} \left[
 \lambda_1 \left( N + \frac{Q^2}{ N} \right)
+ \lambda_2 (  N- \frac{Q^2}{ N}) \right] }
\label{numepqsphdiagtzero}
\end{eqnarray}
The leading behavior of the denominator of Eq. \ref{denopq} reads using Eq. \ref{zsphzeroT}
\begin{eqnarray}
 {\cal Z}_{\cal J}(\beta)  = (Z_{\cal J}^{single}(\beta) )^2 \propto e^{ \beta N  \lambda_1    }
\label{denopqsph}
\end{eqnarray}
so the overlap distribution of Eq. \ref{defpq} reads near zero-temperature
\begin{eqnarray}
{\cal P}_{\cal J}(Q) \propto \frac{{\cal Z}_{\cal J}(\beta ;  Q)}{(Z_{\cal J}(\beta))^2} && \propto   
  e^{ - \frac{\beta}{2} N   (\lambda_1-\lambda_2) \left(  1 - \frac{Q^2}{ N^2} \right) }
\label{pqsphe}
\end{eqnarray}
i.e. in the rescaled variable $q=Q/N$
\begin{eqnarray}
P_{\cal J}(q) = N {\cal P}_{\cal J}(Q=Nq)
\propto  e^{ - \frac{\beta}{2} N   (\lambda_1-\lambda_2) \left(  1 - q^2 \right) }
\label{pqsherescalfin}
\end{eqnarray}
The difference between the two largest eigenvalues reads \cite{dieng,forrester}
\begin{eqnarray}
\lambda_1-\lambda_2 = \frac{v}{N^{2/3}}
\label{difflambda12}
\end{eqnarray}
where $v$ is a positive random variable of order $O(1)$, whose distribution can be obtained
from the joint distribution of $(\lambda_1,\lambda_2)$ \cite{dieng} (here we need the GOE case, but see \cite{forrester} for the neighboring case of GUE matrices).
Plugging Eq. \ref{difflambda12} yields the final result
\begin{eqnarray}
P_{\cal J}(q)  && \propto   
  e^{ -\frac{\beta}{2}  N^{1/3}   \left(  1 - q^2 \right) v }
\label{pqsherescal}
\end{eqnarray}
In particular, 
the typical value decays exponentially in $N^{\theta}=N^{1/3}$ in the whole central region $-1<q<1$
\begin{eqnarray}
\ln P^{typ}(q) \equiv \overline{ \ln P_{\cal J}(q) } \propto  -  \frac{\beta}{2} N^{1/3}   \left(  1 - q^2 \right) \overline{v}
\label{pqsherescaltyp}
\end{eqnarray}
and the appropriate rescaled variable is
\begin{eqnarray}
v=\left( - \frac{ \ln P_{\cal J}(q) } {\frac{\beta}{2}  N^{1/3}   \left(  1 - q^2 \right) } \right) 
\label{pqsherescaldistri}
\end{eqnarray}
which is the $O(1)$ positive random variable of Eq. \ref{difflambda12} for GOE matrices.
In the Gaussian random matrix ensembles, it is well known that there exists a level-repulsion 
between nearest-neighbors eigenvalues as a consequence of the delocalized character of eigenstates,
with the following power-law for the distribution $P(v)$ of the variable $v$ of Eq. \ref{difflambda12}
near the origin $v \to 0$
\begin{eqnarray}
P(v) \oppropto_{v \to 0} v^{a}
\label{vsph}
\end{eqnarray}
where $a=1$ for GOE ($a=2$ for GUE).
This is different from the finite weight $P(v=0)>0$ expected in short-ranged spin-glass models.
As a consequence, the power-law decay of the averaged value in the spherical model
\begin{eqnarray}
P^{av}(q) \equiv \overline{ P_{\cal J}(q) } \propto N^{-x_{sph}}
\label{probenavsph}
\end{eqnarray}
will be different from the simple value of Eq. \ref{xsimple}, and should be instead
\begin{eqnarray}
x_{sph}=(1+a) \theta = \frac{2}{3}
\label{xsph}
\end{eqnarray}

\subsection{ Cumulative overlap distribution in each sample } 

In each sample ${\cal J}$, the cumulative overlap distribution 
will inherit from Eq. \ref{pqsherescalfin} the same exponential decay 
with respect to the size
\begin{eqnarray}
I_{\cal J}(q_0) && \equiv \int_{0}^{q_0} dq P_{\cal J}(q) 
\propto  e^{ - \frac{\beta}{2} N   (\lambda_1-\lambda_2) \left(  1 - q_0^2 \right) }
=  e^{ -\frac{\beta}{2}  N^{1/3}   \left(  1 - q_0^2 \right) v }
\label{cumulativesph}
\end{eqnarray}
where $v$ is the positive random variable of order $O(1)$ of Eq. \ref{difflambda12}.

 \section{ One-dimensional Long-Ranged Ising Spin-Glass  } 

\label{sec_LRSG}

\subsection {Model} 

The one-dimensional long-ranged Ising Spin-Glass \cite{kotliar} is defined by the Hamiltonian
\begin{eqnarray}
 H_{\cal J} && = - \sum_{1 \leq i <j \leq N} J_{ij} S_i S_j
\label{HSGring}
\end{eqnarray}
where the $N$ spins $S_i=\pm$ lie equidistantly on a ring, so that the distance between the
two spins $S_i$ and $S_j$ reads
\begin{eqnarray}
r_{ij}= \frac{N}{\pi} \sin \left(\vert j-i \vert \frac{\pi}{N} \right)
\label{rij}
\end{eqnarray}
The couplings are chosen to decay with some power-law of this distance
\begin{eqnarray}
J_{ij}= c_N(\sigma) \frac{\epsilon_{ij}}{r_{ij}^{\sigma}}
\label{defjij}
\end{eqnarray}
where $\epsilon_{ij}$ are random Gaussian variables
of zero mean $\overline{\epsilon}=0$ and unit variance $\overline{\epsilon^2}=1 $.
The constant $c_N(\sigma) $ is defined by the condition
\begin{eqnarray}
1= \sum_{j \ne 1} \overline{J_{1j}^2} =  c_N^2(\sigma) \sum_{j \ne 1} \frac{1}{r_{1j}^{2 \sigma}}
\label{defcsigma}
\end{eqnarray}
It is important to distinguish the two regimes :

(i) For $0 \leq \sigma < 1/2$, there is an explicit size-rescaling of the couplings
\begin{eqnarray}
 c_N(\sigma) \propto N^{\sigma- \frac{1}{2}}
\label{rescalcsigma}
\end{eqnarray}
as in the Sherrington-Kirkpatrick mean-field model that corresponds to the case $\sigma=0$.

(ii) For $\sigma>1/2$, there is no size rescaling of the couplings
\begin{eqnarray}
 c_N(\sigma) =O(1)
\label{norescalcsigma}
\end{eqnarray}
The limit $\sigma =+\infty$ corresponds to the short-ranged one-dimensional model.
There exists a spin-glass phase at low temperature for $\sigma<1$ \cite{kotliar}. 
The critical point  is mean-field-like for $\sigma<2/3$, and non-mean-field-like for $2/3<\sigma<1$ \cite{kotliar}.

In summary, this model allows to interpolate continuously between the one-dimensional short-ranged model $(\sigma=+\infty)$ and the Sherrington-Kirkpatrick mean-field model ( $\sigma=0$), and is much simpler to study numerically than hypercubic lattices as a function of the dimension $d$. This is why this model has 
  attracted a lot of interest recently \cite{KY,KYgeom,KKLH,KKLJH,Katz,almeidaLR,LRmoore,KHY,KH}
(there also exists a diluted version of the model \cite{diluted}).

\subsection {Measure of the droplet exponent $\theta(\sigma)$ }

Since we wished to evaluate minimal excitation-energies such as Eq. \ref{eminp}, we have
chosen to work, in each disordered sample, by exact enumeration of the $2^N$
spin configurations for small sizes $6 \leq N \leq 24$. The
statistics over samples have been obtained
for instance with the following numbers $n_s(N)$ of disordered samples
\begin{eqnarray}
n_s(L \leq 12) = 2\times10^8 ; ...;n_s(L = 16)=10^7 ; ... n_s(L = 22)=10^5 ;  n_s(L = 24)=2\times10^4 
\label{ns1copy}
\end{eqnarray}

\subsubsection{ The droplet exponent as a stiffness exponent }

The droplet exponent $\theta(\sigma)$ as a function of $\sigma$ has been measured via
Monte-Carlo simulations on sizes $L \leq 256$ in \cite{KY}
from the difference of the ground-state energy
between Periodic and Antiperiodic Boundary conditions in each sample (see the Appendix around Eq. \ref{edw}
for more explanations)
\begin{eqnarray}
  E^{GS(P)}_{\cal J}-E^{GS(AP)}_{\cal J}  = N^{\theta} u
\label{edw1d}
\end{eqnarray}
where $u$ is an $O(1)$ random variable of zero mean (with a probability distribution
symmetric in $u \to -u$).
In this context, 'Antiperiodic' means the following prescription \cite{KY} :
for each disordered sample $(J_{ij})$ considered as 'Periodic', the 'Antiperiodic'
consists in changing the sign $J_{ij} \to -J_{ij}$ for all pairs $(i,j)$ where
the shortest path on the circle goes through the bond $(L,1)$.
We have followed exactly the same procedure, and our results via exact enumeration on much
smaller sizes $6 \leq L \leq 24$
 for the three values of $\sigma$ we have considered, are actually close to the values given in \cite{KY}
\begin{eqnarray}
\theta(\sigma=0.1) \simeq 0.3
\nonumber \\
\theta(\sigma=0.62) \simeq 0.24
\nonumber \\
\theta(\sigma=0.75) \simeq 0.17
\label{thetadw1d}
\end{eqnarray}
We refer the reader to Ref \cite{KY} for other values of $\sigma$.

\subsubsection{ Statistics over samples of the ground state energy  }

We have also studied the statistics of the ground state energy over samples (see the Appendix around Eq
\ref{e0av} and \ref{e0fluct} for more explanations).
We find that the correction to extensivity of the averaged ground state energy (see Eq \ref{e0av})
\begin{eqnarray}
\overline{ E_{\cal J}^{GS}(N) }  \simeq N e_0+ N^{\theta_{shift}} e_1+...
\label{e0av1d}
\end{eqnarray}
is governed by the droplet exponent measured in Eq. \ref{thetadw1d} from Eq. \ref{edw1d}
\begin{eqnarray}
\theta_{shift}(\sigma)= \theta(\sigma)
\label{thetashift1d}
\end{eqnarray}
as expected in general within the droplet scaling theory (Eq \ref{thetashift}).

We have also measured 
the fluctuation exponent $\mu$ of Eq \ref{e0fluct}
\begin{eqnarray}
\mu(\sigma=0.1) \simeq 0.3 
\nonumber \\
\mu(\sigma=0.62) \simeq 0.35
\nonumber \\
\mu(\sigma=0.75) \simeq 0.4 
\label{mu1d}
\end{eqnarray}
The last two values are in agreement with Ref \cite{KY}
whereas the first value is larger than the value $\mu(\sigma=0.1) \simeq 0.25 $  
of Ref. \cite{KY}. 
We refer the reader to Ref \cite{KY} for other values of $\sigma$.

\begin{figure}[htbp]
\includegraphics[height=6cm]{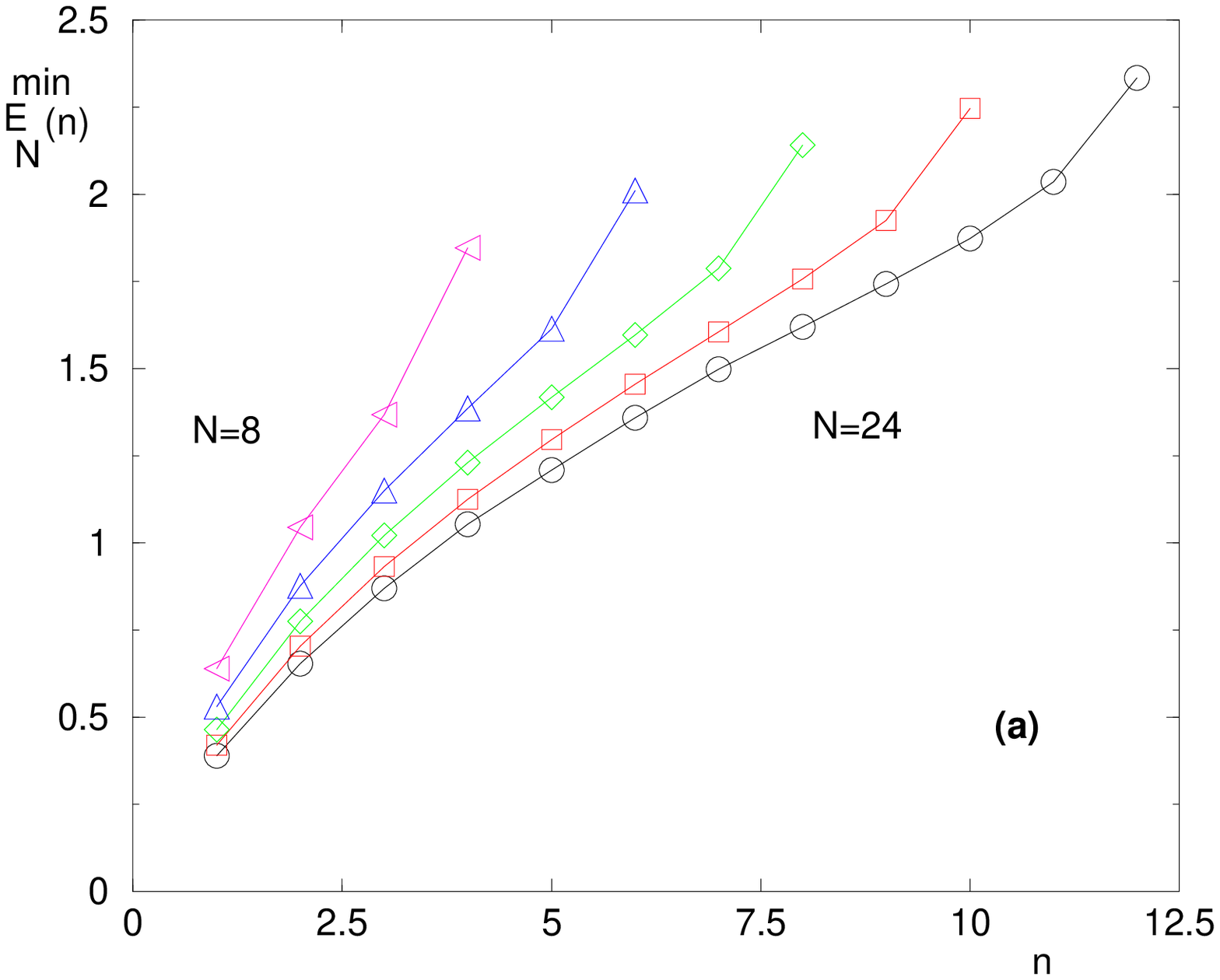}
\hspace{1cm}
 \includegraphics[height=6cm]{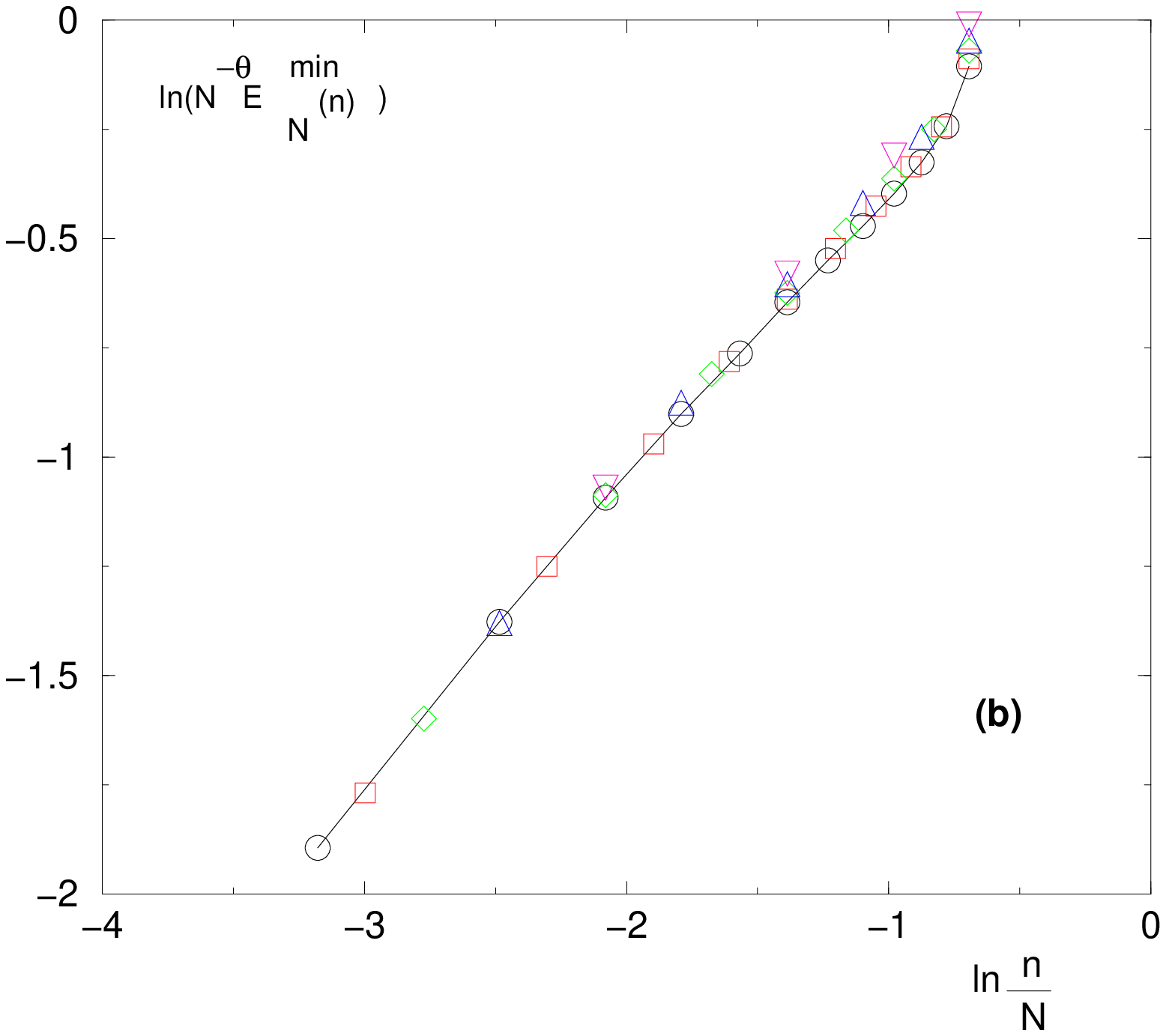}
\caption{ Statistics of the minimal excitation energy involving $n$ spins (Eq \ref{eminp1d}) in the one-dimensional long-ranged model 
of power $\sigma=0.1$  :
(a) Average over samples $E_{N}^{min}(n) \equiv  \overline{ E_{\cal J}^{min}(n) }$ as a function of $n$ for sizes $N=8,12,16,20, 24$
(b) Data-collapse obtained by testing the scaling form of Eq. \ref{eminp1drescal} :  
$\ln (N^{-\theta} E_{N}^{min}(n))$ as a function of $\ln \frac{n}{N} $ with $\theta \simeq 0.3$
 }
\label{figeminsigma=0.1}
\end{figure}

\begin{figure}[htbp]
\includegraphics[height=6cm]{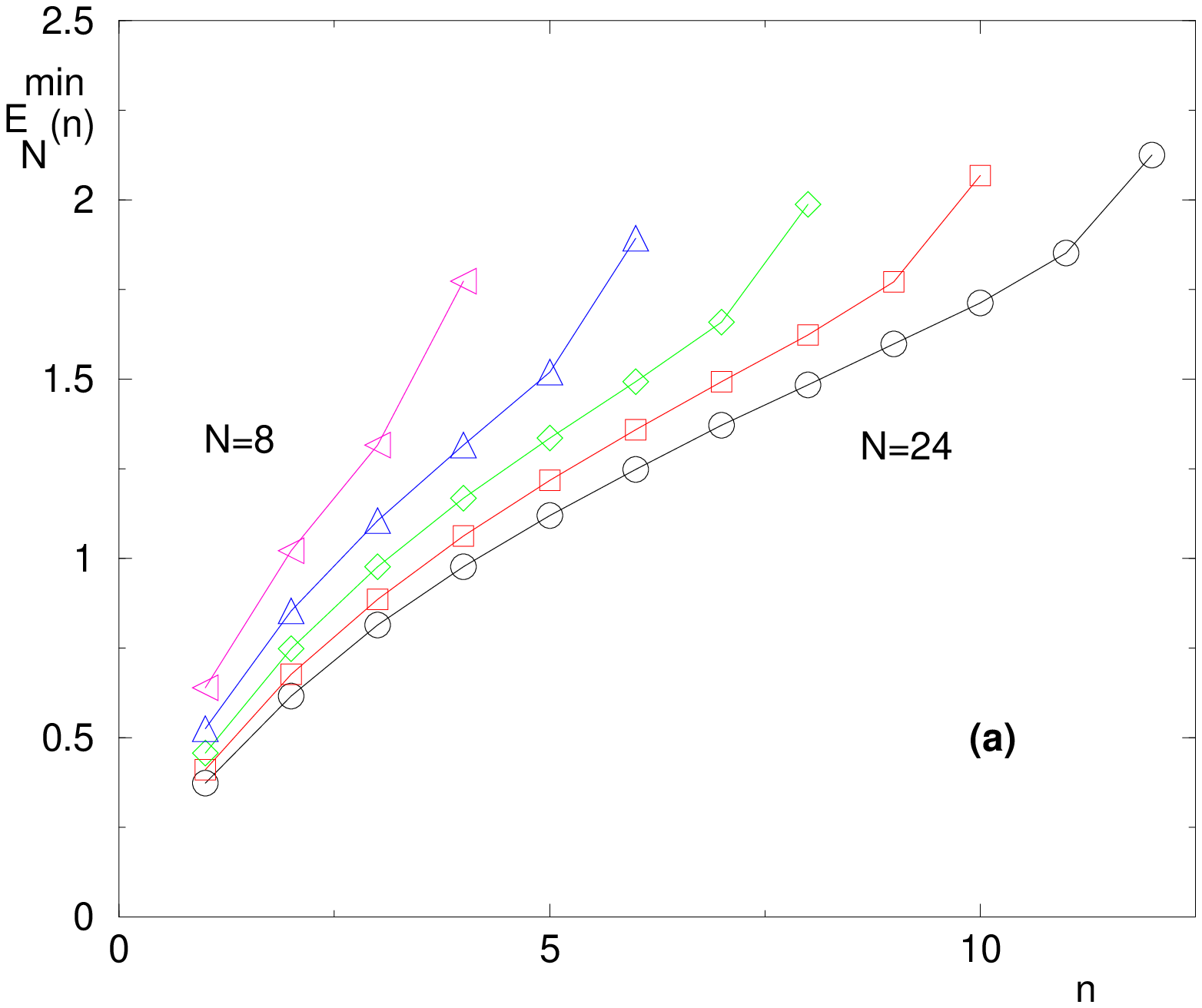}
\hspace{1cm} 
\includegraphics[height=6cm]{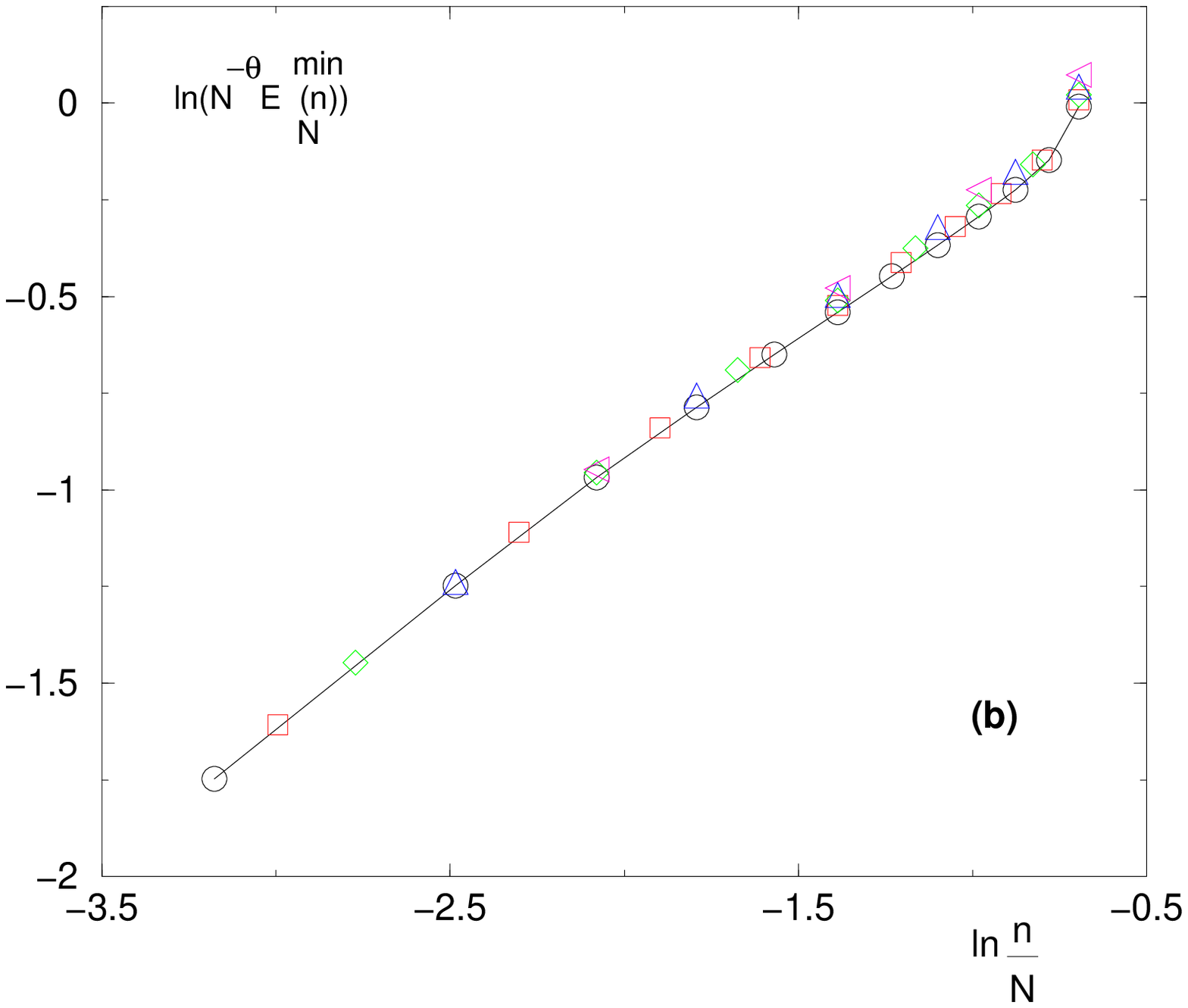}
\caption{ Statistics of the minimal excitation energy involving $n$ spins (Eq \ref{eminp1d}) in the one-dimensional long-ranged model 
of power $\sigma=0.62$  :
(a) Average over samples $E_{N}^{min}(n) \equiv  \overline{ E_{\cal J}^{min}(n) }$ as a function of $n$ for sizes $N=8,12,16,20, 24$
(b) Data-collapse obtained by testing the scaling form of Eq. \ref{eminp1drescal} :  
$\ln (N^{-\theta} E_{N}^{min}(n))$ as a function of $\ln \frac{n}{N} $ with $\theta \simeq 0.24$
 }
\label{figeminsigma=0.62}
\end{figure}

\begin{figure}[htbp]
\includegraphics[height=6cm]{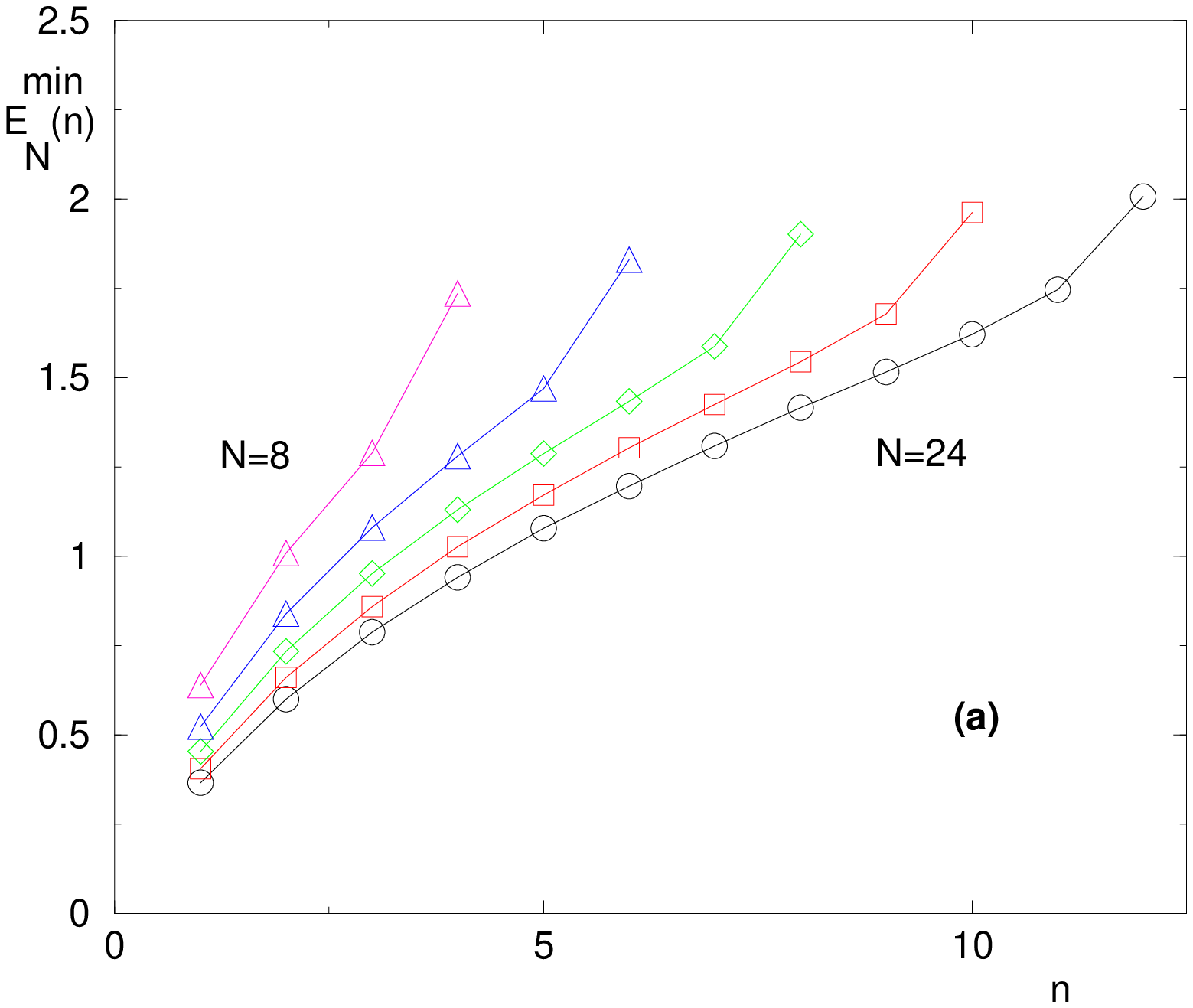}
\hspace{1cm}
 \includegraphics[height=6cm]{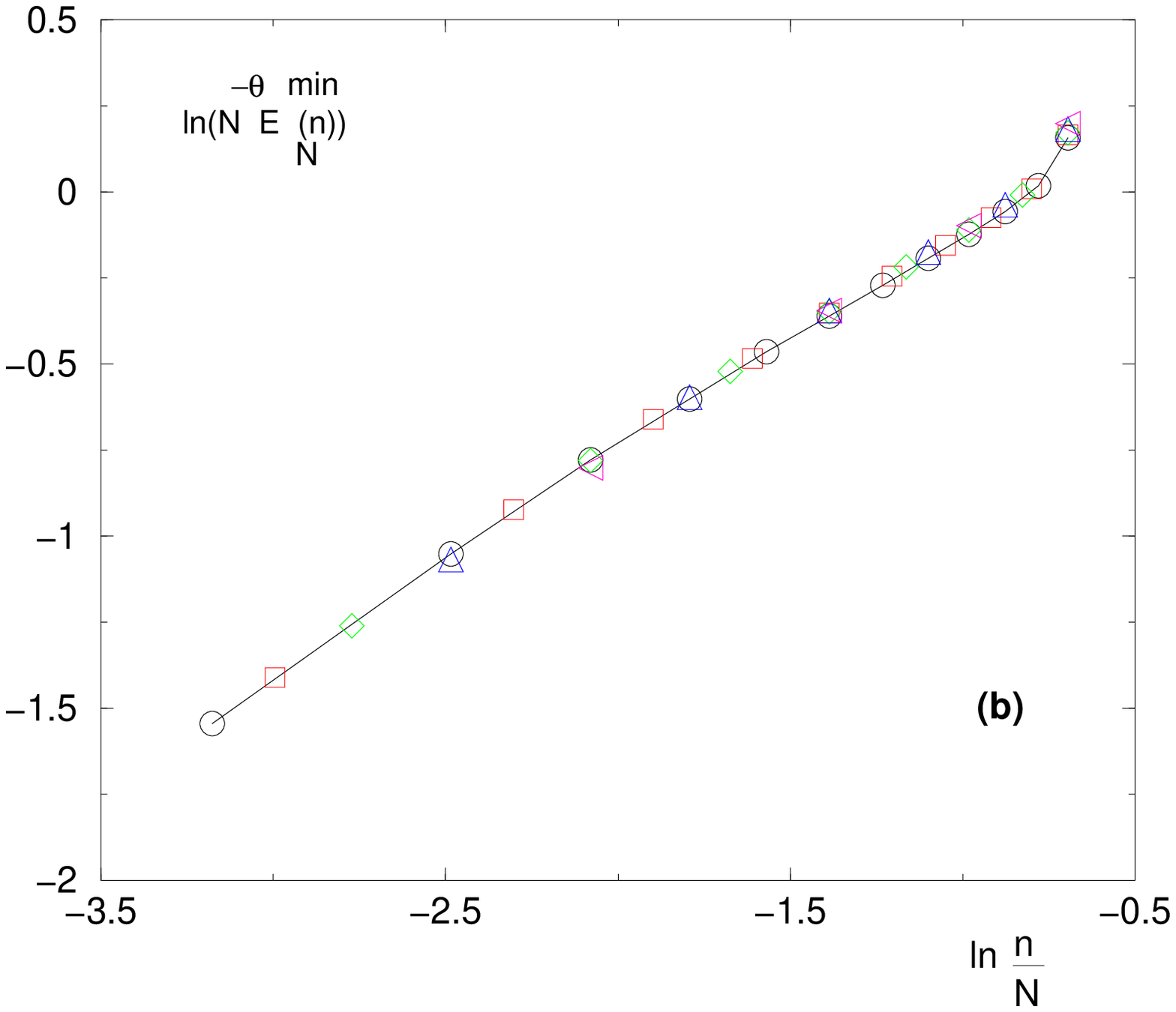}
\caption{Statistics of the minimal excitation energy involving $n$ spins (Eq \ref{eminp1d}) in the one-dimensional long-ranged model 
of power $\sigma=0.75$  :
(a) Average over samples $E_{N}^{min}(n) \equiv  \overline{ E_{\cal J}^{min}(n) }$ as a function of $n$ for sizes $N=8,12,16,20, 24$
(b) Data-collapse obtained by testing the scaling form of Eq. \ref{eminp1drescal} :  
$\ln (N^{-\theta} E_{N}^{min}(n))$ as a function of $\ln \frac{n}{N} $ with $\theta \simeq 0.17$
 }
\label{figeminsigma=0.75}
\end{figure}

\subsubsection{ Minimal energy of fixed-size excitations  in a given sample }

We have measured in each sample ${\cal J}$ the minimal energy cost $(E_{\cal J}(i_1,..,i_n)-E_{GS})$ among all 
excitations involving the flipping of exactly $n$ spins with respect to the ground state (Eq \ref{eminp})
\begin{eqnarray}
E_{\cal J}^{min}(n) \equiv  \min_{1 \leq i_1 <i_2 < .. < i_n \leq N} \left(E_{\cal J}(i_1,..,i_n)-E_{\cal J}^{GS}\right) 
\label{eminp1d}
\end{eqnarray}
We show on Figures \ref{figeminsigma=0.1}, \ref{figeminsigma=0.62}, \ref{figeminsigma=0.75}
that our data for the averaged value over the samples ${\cal J}$ of size $N$
\begin{eqnarray}
E_{N}^{min}(n) \equiv  \overline{ E_{\cal J}^{min}(n) }
\label{eminp1dav}
\end{eqnarray}
can be rescaled in the following form
\begin{eqnarray}
E_{N}^{min}(n) \simeq  N^{\theta(\sigma)} g \left(  \frac{n}{N} \right)
\label{eminp1drescal}
\end{eqnarray}
where $\theta(\sigma)$ is the droplet exponent measured previously in Eq. \ref{thetadw1d}.

Since we expect that this averaged value governs the low-temperature behavior of the typical overlap
 (Eq \ref{pqtnearzeromintyp}), the scaling form of Eq. \ref{eminp1drescal}
corresponds to the expectation of Eq. \ref{defpqrescaltheta}.

\subsection { Typical versus  averaged overlap distributions  }

\begin{figure}[htbp]
 \includegraphics[height=6cm]{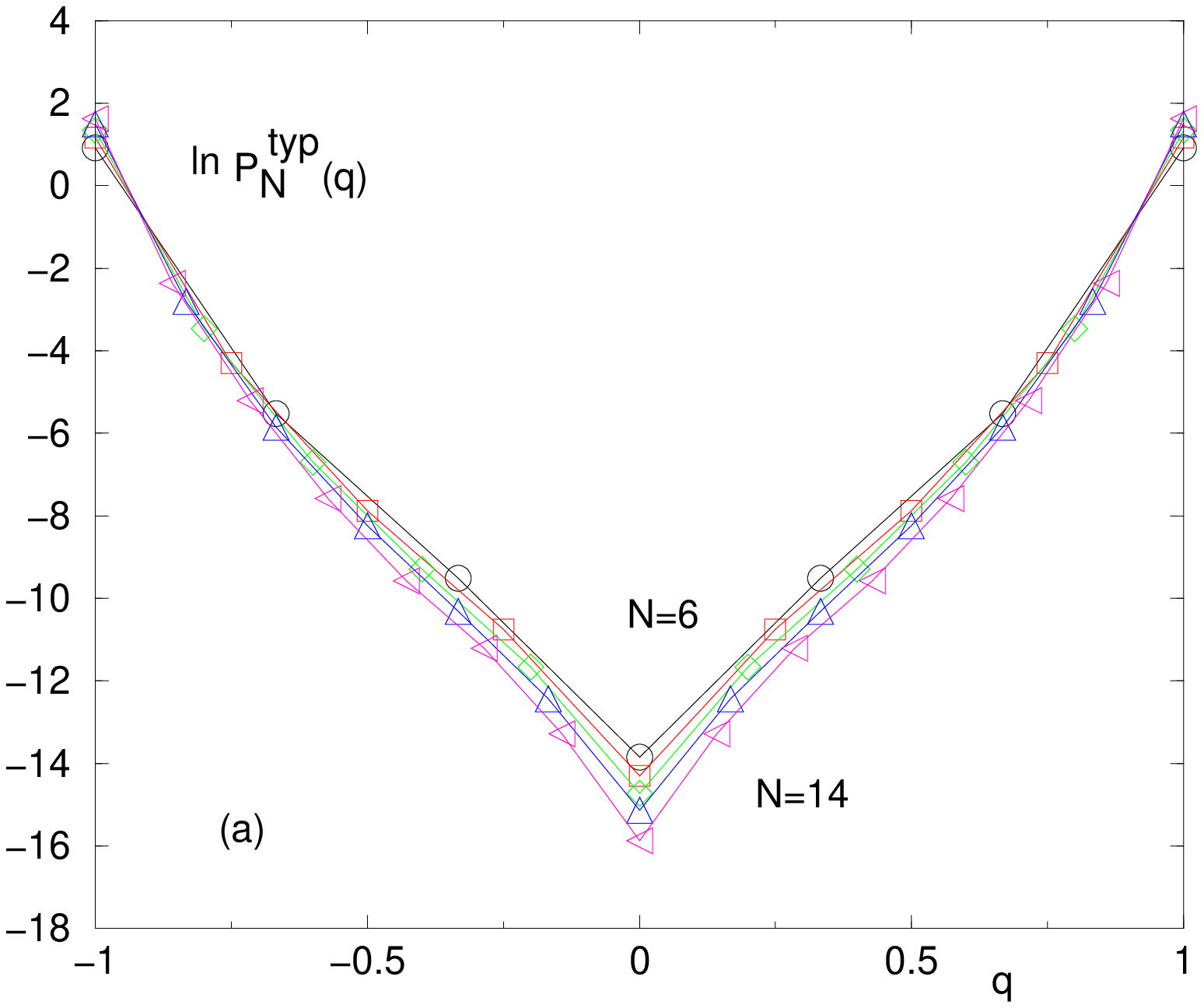}
 \hspace{1cm}
 \includegraphics[height=6cm]{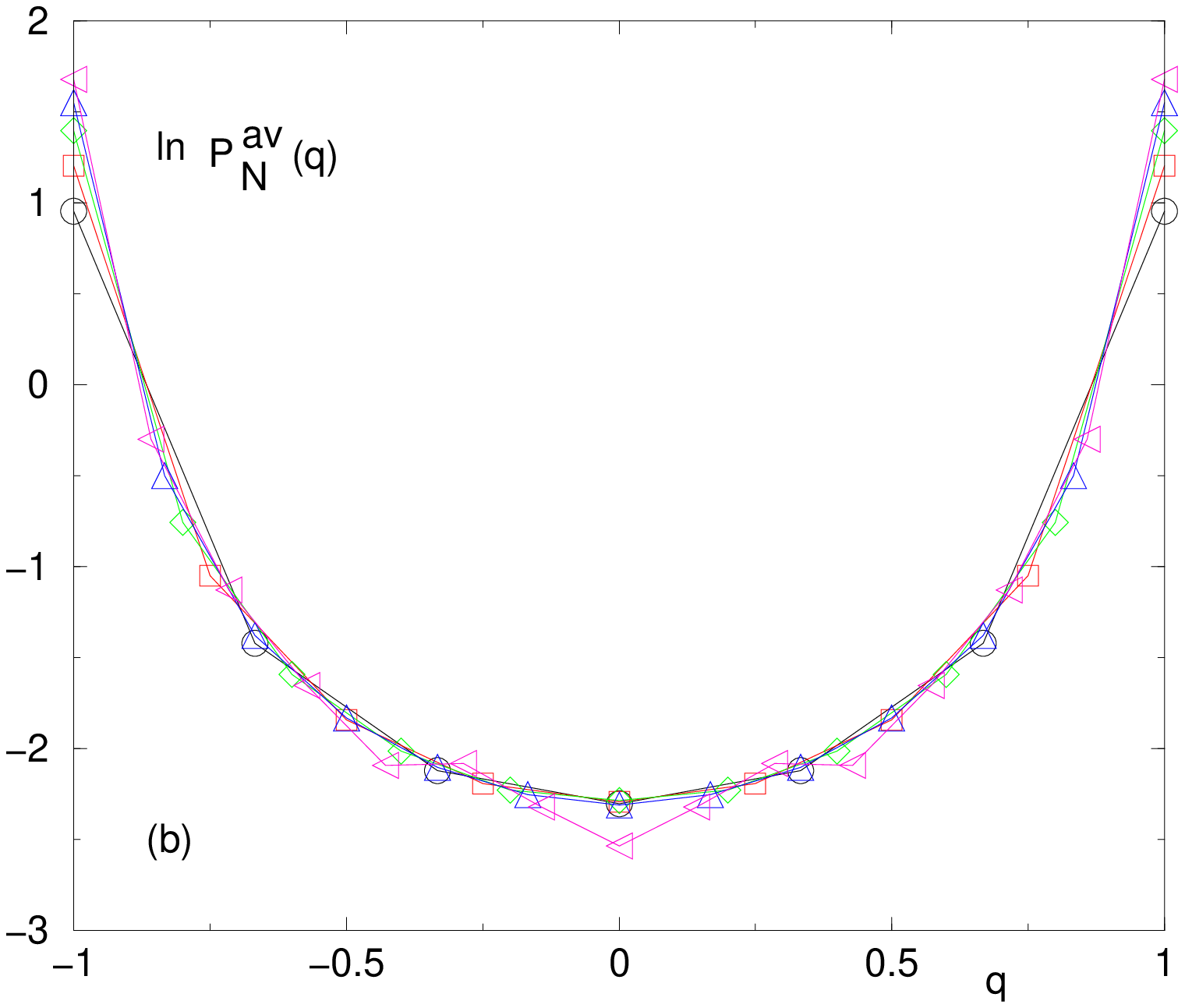}
\caption{ Typical versus averaged overlap distribution for the one-dimensional long-ranged model 
of power-law $\sigma=0.1$ at temperature $T=0.1$ :
(a) $\ln P_N^{typ}(q)$ as a function of $q$  for the sizes $N=6,8,10,12,14$
(a) $\ln P_N^{av}(q)$ as a function of $q$  for the sizes $N=6,8,10,12,14$  }
\label{figsigma=0.1}
\end{figure}

\begin{figure}[htbp]
 \includegraphics[height=6cm]{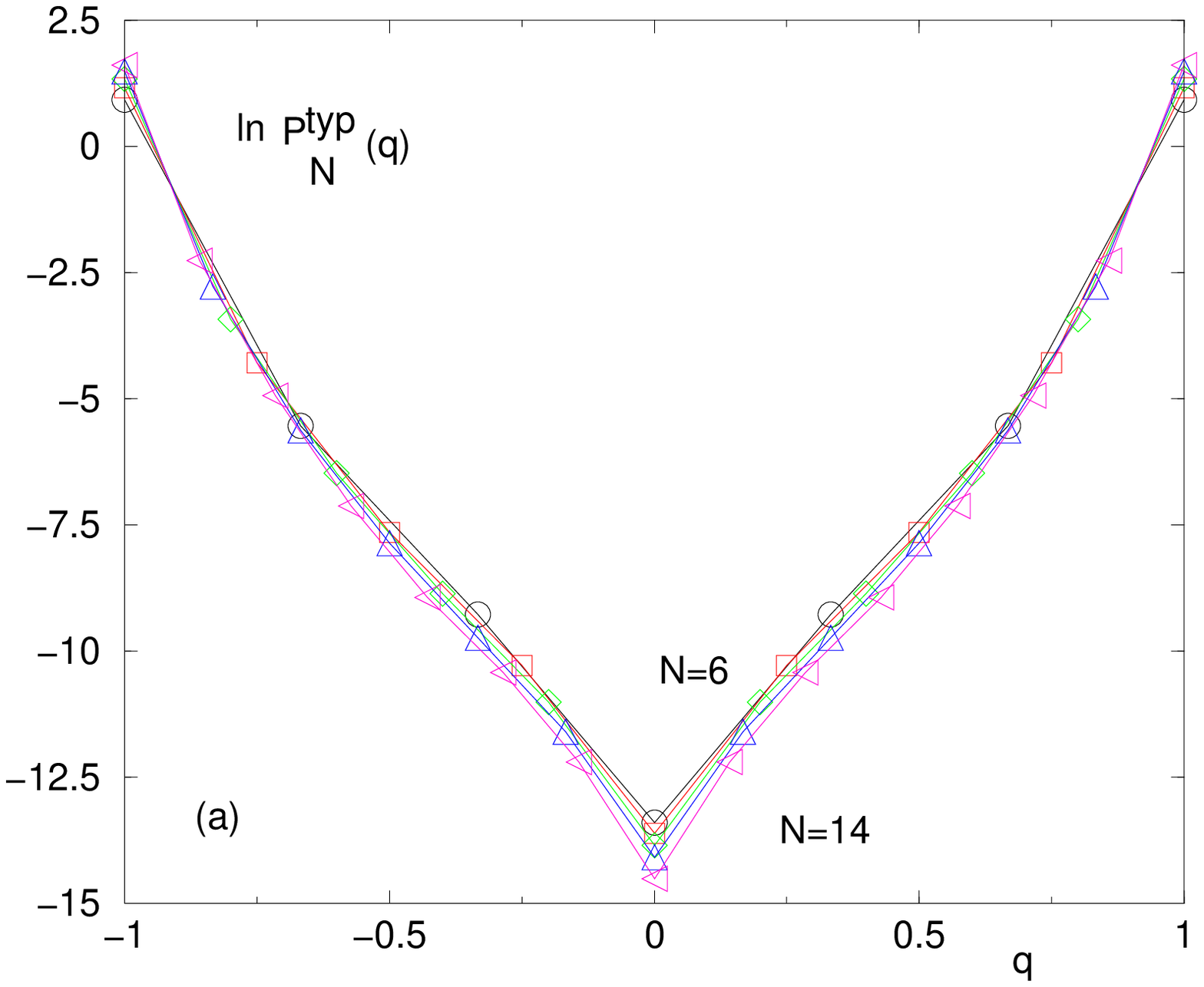}
 \hspace{1cm}
 \includegraphics[height=6cm]{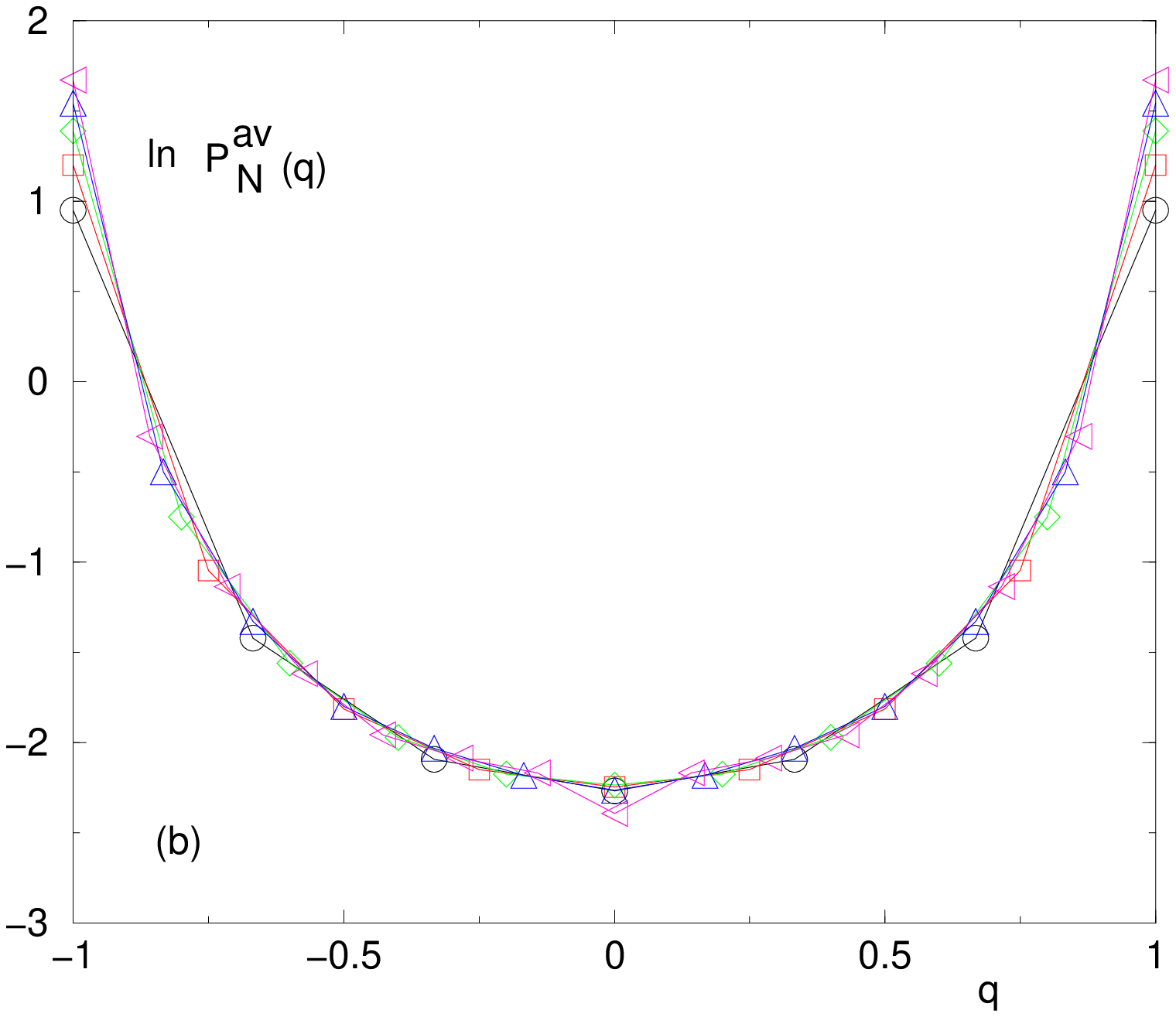}
\caption{ Typical versus averaged overlap distribution for the one-dimensional long-ranged model 
of power-law $\sigma=0.62$ at temperature $T=0.1$ :
(a) $\ln P_N^{typ}(q)$ as a function of $q$  for the sizes $N=6,8,10,12,14$
(a) $\ln P_N^{av}(q)$ as a function of $q$  for the sizes $N=6,8,10,12,14$  }
\label{figsigma=0.62}
\end{figure}

\begin{figure}[htbp]
 \includegraphics[height=6cm]{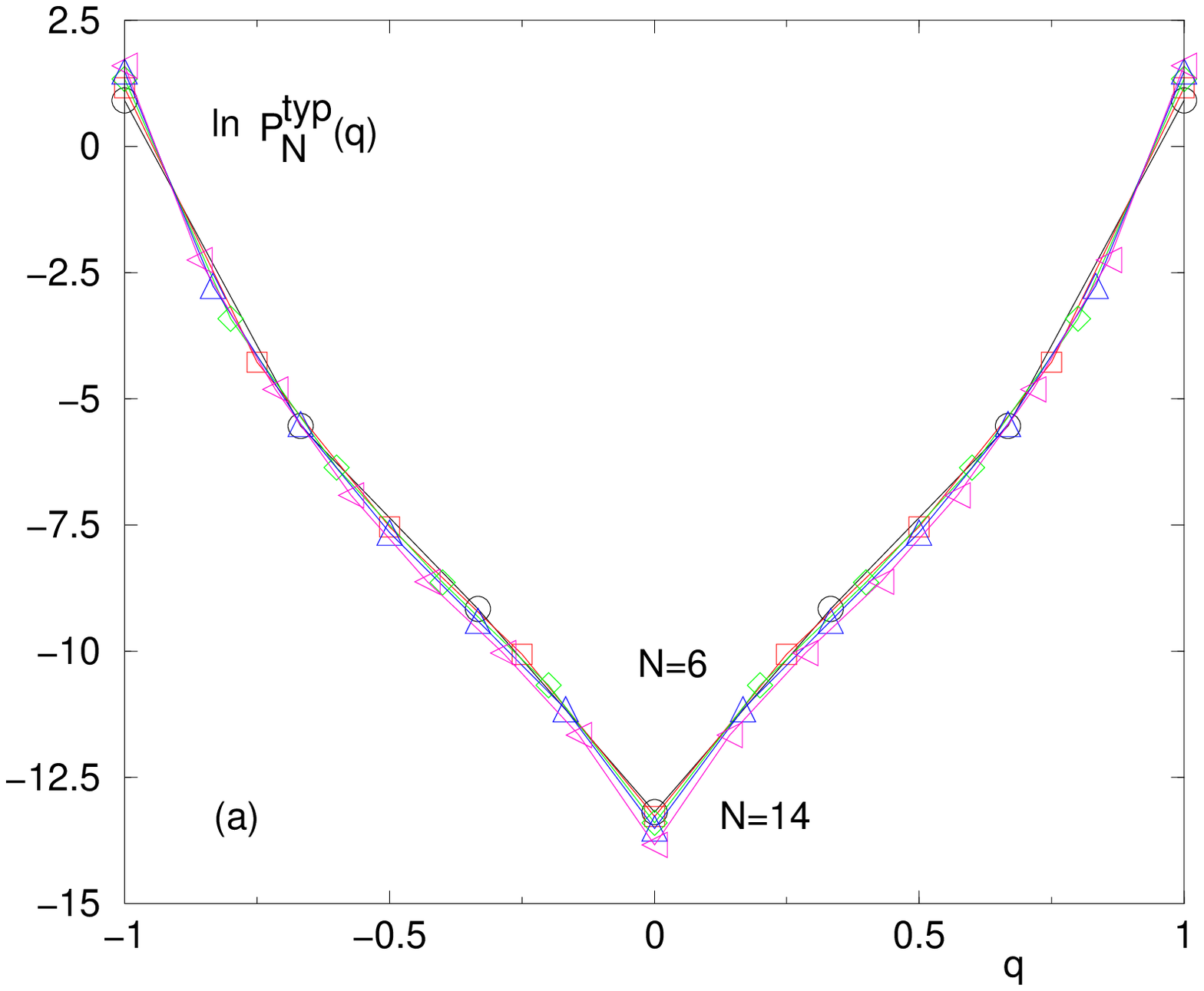}
 \hspace{1cm}
 \includegraphics[height=6cm]{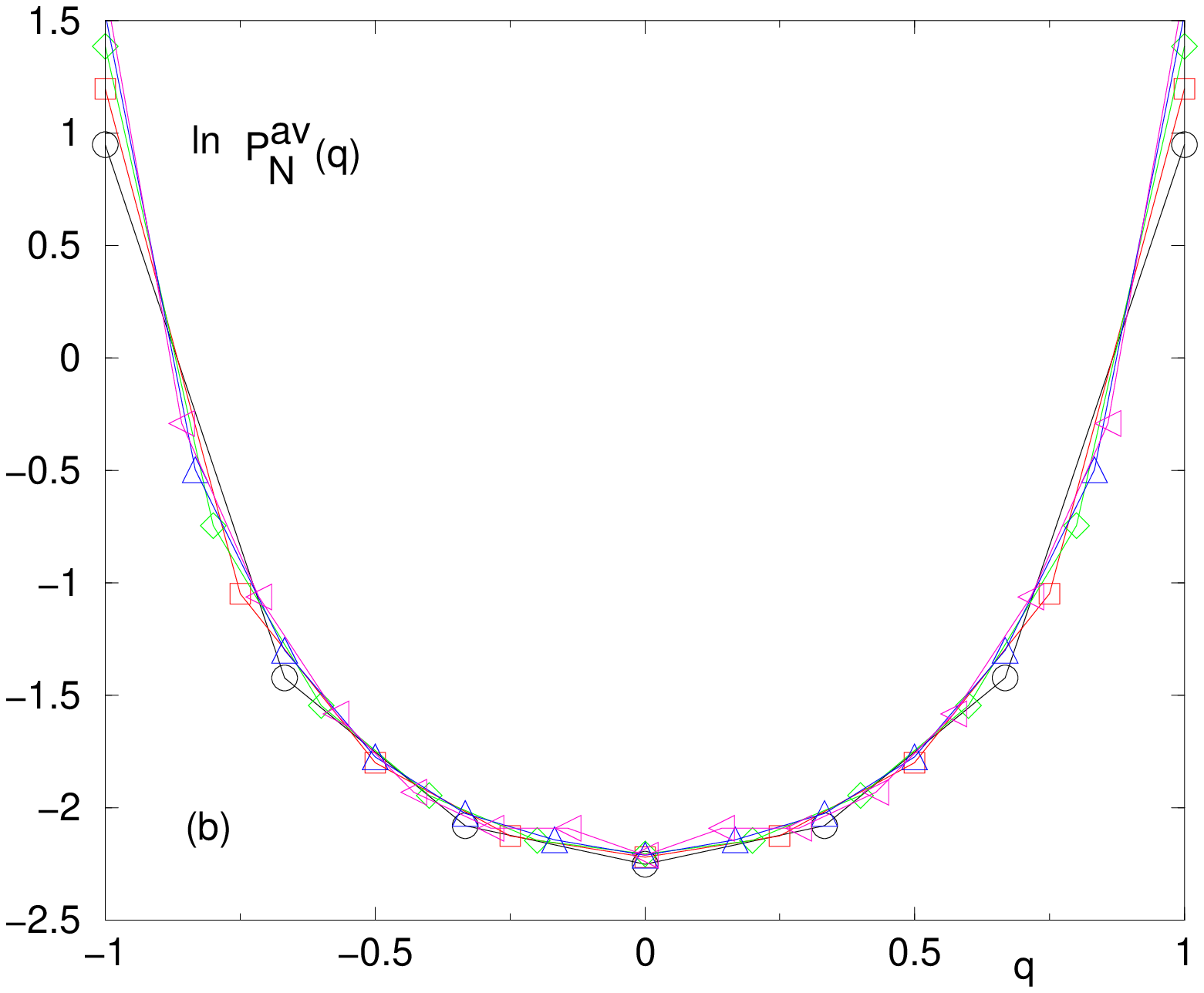}
\caption{ Typical versus averaged overlap distribution for the one-dimensional long-ranged model 
of power-law $\sigma=0.75$ at temperature $T=0.1$ :
(a) $\ln P_N^{typ}(q)$ as a function of $q$  for the sizes $N=6,8,10,12,14$
(a) $\ln P_N^{av}(q)$ as a function of $q$  for the sizes $N=6,8,10,12,14$  }
\label{figsigma=0.75}
\end{figure}

We have also computed directly the overlap distribution $P_{\cal J} (q)$ via exact enumeration of the
$4^N$ configurations of the two copies of spins
for the sizes $6 \leq N \leq 15$ at the temperature $T=0.1$
with the following statistics for the number $n_S(L)$ of samples 
\begin{eqnarray}
n_s(L = 6) = 2 \times 10^8 ; n_s(L = 8) = 2 \times 10^7 ;n_s(L = 10)=10^6 ; n_s(L = 12)=5 \times 10^4 ;  n_s(L = 14)=1750 
\label{ns2copy}
\end{eqnarray}

On Figures \ref{figsigma=0.1},  \ref{figsigma=0.62}, and \ref{figsigma=0.75}, we compare the typical and the averaged 
overlap distribution for three values of the power $\sigma$ : in all cases, we find that they are completely different, in order of magnitudes (see the differences in log-scales)
and in dependence with the system size $N$ : whereas the averaged value does not change rapidly with $N$ (as found also on bigger sizes \cite{KY}),
the typical overlap distribution decays with $N$ in the central region around $q=0$.
 This effect should be even clearer with the large system-sizes used in Ref \cite{KY}.

 \section{ Fully connected Sherrington-Kirkpatrick model   } 

\label{sec_SK}

The fully connected Sherrington-Kirkpatrick Ising spin-glass model \cite{SKmodel}
\begin{eqnarray}
 H_{\cal J} = - \sum_{1 \leq i <j \leq N} J_{ij} S_i S_j
\label{defSK}
\end{eqnarray}
where the couplings $J_{ij}$ are random quenched variables
of zero mean $\overline{J}=0$ and of variance $\overline{J^2}=1/N $ can be seen as the limit
$\sigma=0$
of the one-dimensional long-ranged modem described in the previous section.

\subsection{ Statistics of the ground state}

The statistics over samples of the ground state energy  has been much studied
in the SK model
\cite{andreanov,Bou_Krz_Mar,pala_gs,aspelmeier_MY,Katz_gs,Katz_guiding,aspelmeier_BMM,boettcher_gs,us_tails,us_matching}.
There seems to be a consensus on the shift exponent governing the correction to extensively of
the averaged value (Eq \ref{e0av})
\begin{eqnarray}
 \theta_{shift} \simeq 0.33
\label{shiftSK}
\end{eqnarray}
which is thus close to the value of the long-ranged one dimensional model for $\sigma =0.1$
discussed above. With our exact enumeration on small sizes $6 \leq N \leq 24$, we see the compatible value
\begin{eqnarray}
 \theta \simeq 0.31
\label{thetask}
\end{eqnarray}

On the contrary,  the `fluctuation exponent' $\mu$ is controversial,
with the two main proposals $\mu =1/4$ and $\mu=1/6$
(see the discussions in
\cite{andreanov,Bou_Krz_Mar,pala_gs,aspelmeier_MY,Katz_gs,Katz_guiding,aspelmeier_BMM,boettcher_gs,us_tails,us_matching}).

\subsection{ Minimal energy of fixed-size excitations  in a given sample }

\begin{figure}[htbp]
 \includegraphics[height=6cm]{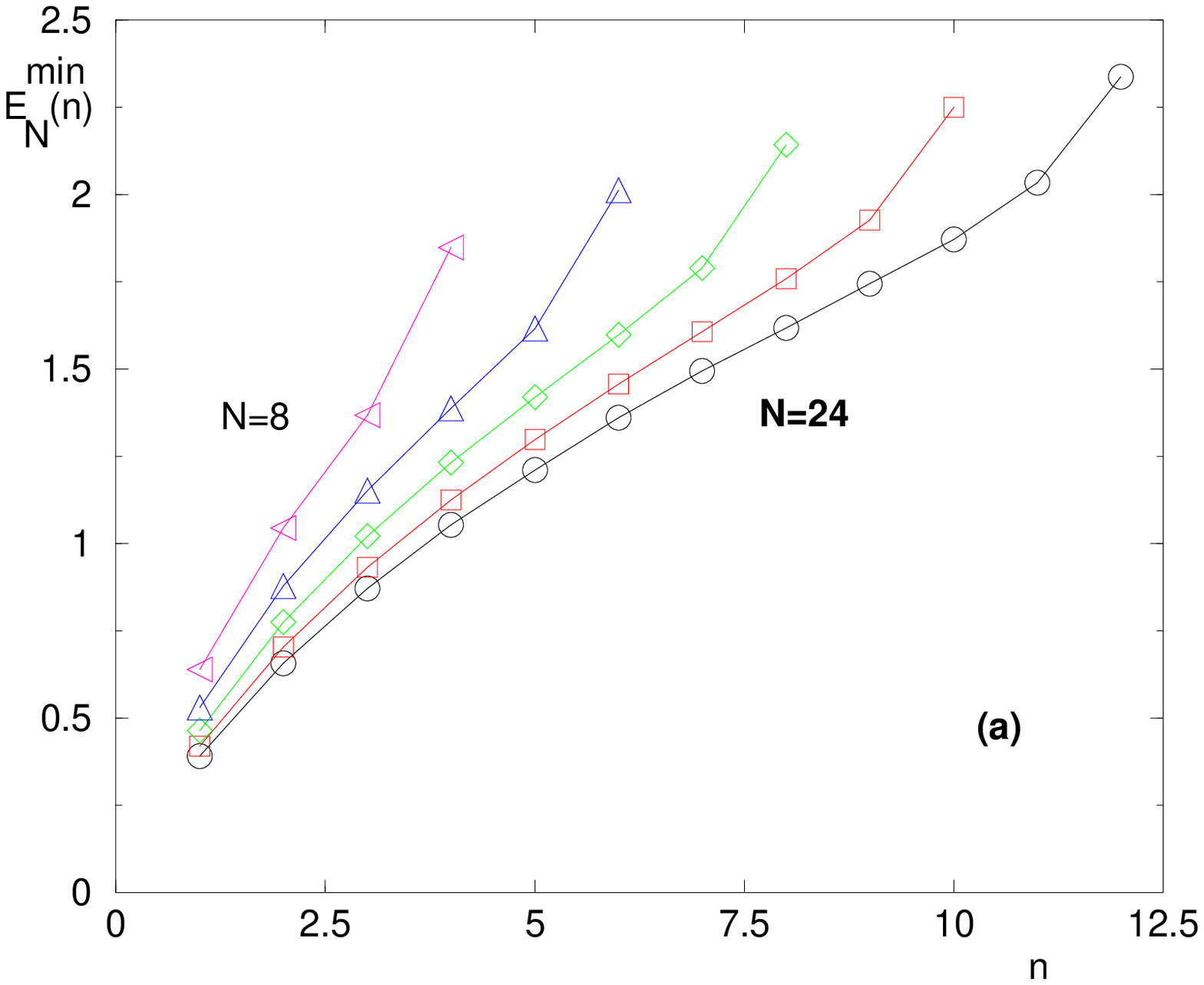}
 \hspace{1cm}
 \includegraphics[height=6cm]{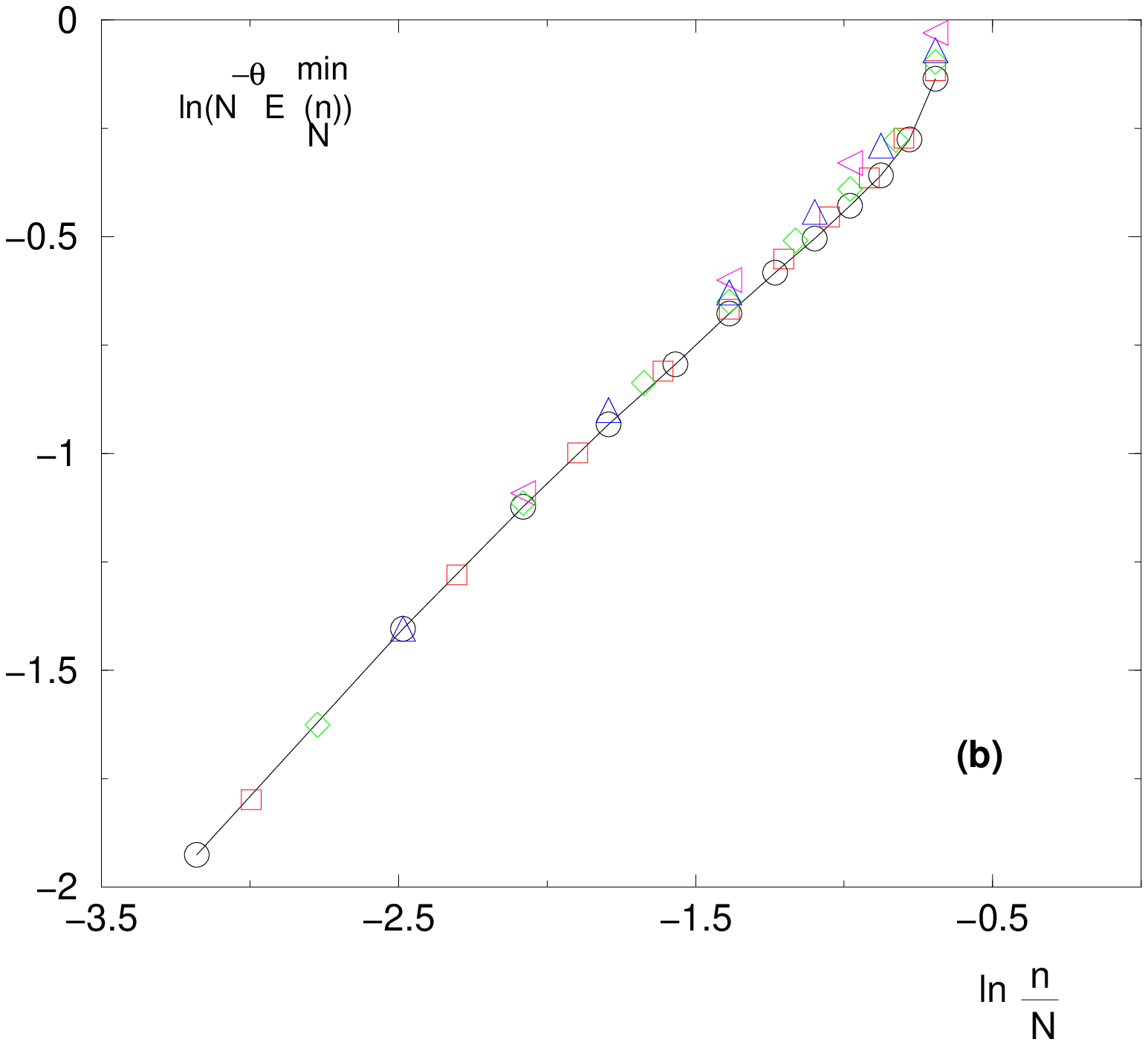}
\caption{Statistics of the minimal excitation energy involving $n$ spins (Eq \ref{eminp1dav}) in the mean-field SK model  :
(a) Average over samples $E_{N}^{min}(n) \equiv  \overline{ E_{\cal J}^{min}(n) }$ as a function of $n$ for sizes $N=8,12,16,20, 24$
(b) Data-collapse obtained by testing the scaling form of Eq. \ref{eminp1drescal} :  
$\ln (N^{-\theta} E_{N}^{min}(n))$ as a function of $\ln \frac{n}{N} $ with $\theta \simeq 0.31$
 }
\label{figeminsk}
\end{figure}

We show on Figures \ref{figeminsk} that our data for the averaged value over the samples ${\cal J}$ of size $N$
of the minimal energy cost $(E_{\cal J}(i_1,..,i_n)-E_{GS})$ among all 
excitations involving the flipping of exactly $n$ spins with respect to the ground state (Eq \ref{eminp})
\begin{eqnarray}
E_{N}^{min}(n) \equiv  \overline{ E_{\cal J}^{min}(n) }
\label{eminp1davsk}
\end{eqnarray}
can be rescaled in the following form
\begin{eqnarray}
E_{N}^{min}(n) \simeq  N^{\theta} g \left(  \frac{n}{N} \right)
\label{eminp1drescalsk}
\end{eqnarray}
with $\theta \simeq 0.31$ is the droplet exponent measured previously in Eq. \ref{thetask}.

Since we expect that this averaged value governs the low-temperature behavior of the typical overlap
 (Eq \ref{pqtnearzeromintyp}), the scaling form of Eq. \ref{eminp1drescal}
corresponds to the expectation of Eq. \ref{defpqrescaltheta}.

\begin{figure}[htbp]
 \includegraphics[height=6cm]{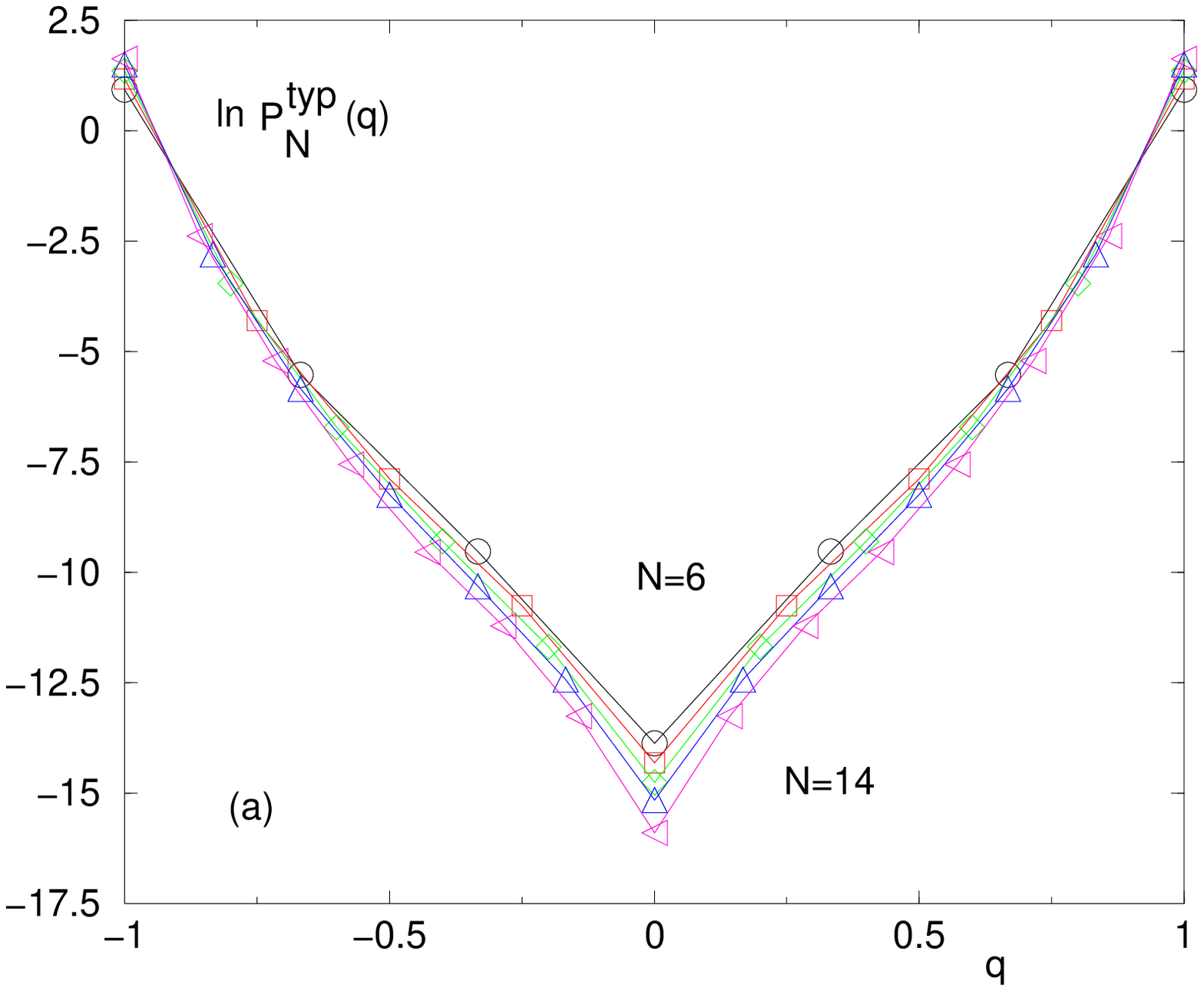}
 \hspace{1cm}
 \includegraphics[height=6cm]{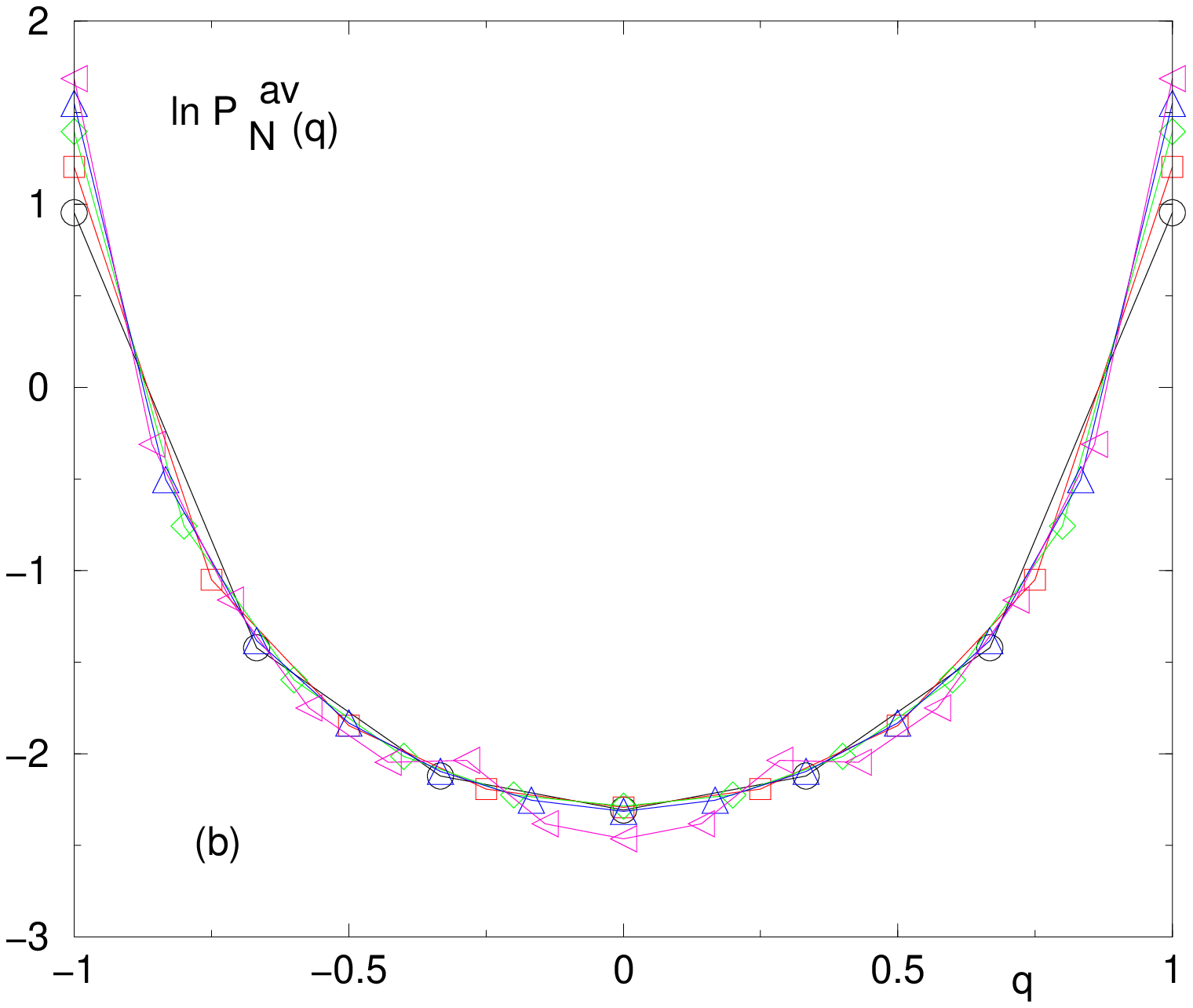}
\caption{ Typical versus averaged overlap distribution for the SK model at temperature $T=0.1$ :
(a) $\ln P_N^{typ}(q)$ as a function of $q$  for the sizes $N=6,8,10,12,14$
(a) $\ln P_N^{av}(q)$ as a function of $q$  for the sizes $N=6,8,10,12,14$  }
\label{figsk}
\end{figure}

\subsection { Typical versus  averaged overlap distribution  }

We have also computed the overlap distribution $P_{\cal J} (q)$ via exact enumeration of the
$4^N$ configurations of the two copies of spins
for the sizes $6 \leq N \leq 14$ at the temperature $T=0.1$,
with the same statistics as in Eq. \ref{ns2copy}.
As shown on Fig \ref{figsk}, we find again that the typical and the averaged 
overlap distribution are completely different, in order of magnitudes (see the differences in log-scales)
and in dependence with the system size $N$ : whereas the averaged value does not change much with $N$,
the typical overlap distribution clearly decays with $N$ in the central region around $q=0$.

 \section{ Conclusion  } 

\label{sec_conclusion}

In this paper, we have studied the statistical properties over disordered samples $(\cal J)$ of the overlap
distribution $P_{\cal J}(q)$ which plays the role of an order parameter in spin-glasses.
We have obtained  that near zero temperature

 (i) the {\it typical } overlap distribution is
exponentially small in the central region of $-1<q<1$ : 
\begin{eqnarray}
\ln P^{typ} (q) \equiv  \overline { \ln P_{\cal J}(q) } \sim  - \beta N^{\theta} \phi(q)
\label{cptyp}
\end{eqnarray}
where $\theta$ is the droplet exponent
defined here with respect to  the total number $N$ of spins (in order to consider also
fully connected models where the notion of length does not exist).

 (ii) the appropriate rescaled variable to describe sample-to sample fluctuations is
\begin{eqnarray}
v = - \frac{\ln P_{\cal J}(q))}{ \beta N^{\theta}}
\label{cptyprescal}
\end{eqnarray}
which remains an $O(1)$ random positive variable.

 (iii) the averaged distribution $\overline{P_{\cal J}(q) } $ is non-typical, dominated by
 rare anomalous samples and can be thus very misleading.

We have first derived these results for the spherical mean-field model with $\theta=1/3$, $\phi(q)=1-q^2 $, and the random
variable $v$ corresponds
 to the rescaled difference between the two largest eigenvalues of
GOE random matrices. 

We have then presented numerical results for the long-ranged
one-dimensional spin-glass with random couplings decaying as $J(r) \propto r^{-\sigma}$
for various values of the exponent $\sigma$,  and for the SK-mean-field model (corresponding formally to the
$\sigma=0$ limit of the previous model). In all cases, we have obtained that the typical and averaged overlap
distributions are completely different, in order of magnitude and in scaling. 
We have also found that in each case, the same droplet exponent governs the four properties we have measured :

 (a) the change in the ground state energy between different boundary conditions in a given sample

(b) the correction to extensivity of the averaged ground-state energy

(c) the minimal energy of excitations of a fixed extensive size $n \propto N$

(d) the decay of the typical overlap distribution

Our results are thus in full agreement with the droplet scaling theory.

We hope that future studies on spin-glasses will also measure the {\it typical} values
of the overlap distribution $P^{typ}(q)= e^{\overline { \ln P_{\cal J}(q) }}$ or of the cumulative overlap distribution $I^{typ}(q_0)= e^{\overline { \ln I_{\cal J}(q_0) }}$
  instead of the non-typical averaged overlap distribution,
in order obtain clearer conclusions on the nature of the spin-glass phase.

\appendix

 \section{ Brief reminder on some properties of the droplet exponent $\theta$  }

\label{app_theta}

In the droplet scaling theory \cite{mcmillan,bray_moore,fisher_huse}, the most important
notion is the droplet exponent $\theta$, with the following physical meanings.

\subsection{ Scaling of renormalized couplings }

The initial meaning of the droplet exponent $\theta_{l}$ is the scaling of renormalized
couplings $J$
on a length scale $L$ \cite{mcmillan,bray_moore}
\begin{eqnarray}
J_L \simeq L^{\theta_l} u
\label{jr}
\end{eqnarray}
where $u$ is an $O(1)$ random variable of zero mean (with a probability distribution
symmetric in $u \to -u$).
This definition is directly used in real-space renormalization studies based on the
Migdal-Kadanoff approximation
\cite{hierarchicalspinglass}.
The definition of Eq. \ref{jr} means that there is no spin-glass phase when $\theta_l<0$,
and that there exists a spin-glass phase when $\theta_l>0$, which is then governed by a zero-temperature fixed point.

 \subsection{ Scaling of 'optimized excitations' around a given point  }

The above definition can also be interpreted as the energy scale of 'optimized'
excitations of linear size $L$
around a given point \cite{fisher_huse}
\begin{eqnarray}
E^{exc}_L \simeq L^{\theta_l} v
\label{exci}
\end{eqnarray}
where $v$ is a positive random variable.
So the energy scale is $L^{\theta_l}$ with a probability $O(1)$, but can also be $O(1)$
with the small probability $P(v< L^{- \theta_l})$,
so that these rare events will lead to power-laws in various observables
\cite{fisher_huse}.

In the present paper, in order to compare with fully-connected model where the notion of
length does not exist, we have chosen to define the droplet exponent with respect to the
number of spins $N$ involved.
so that Eq. \ref{exci} becomes
\begin{eqnarray}
E^{exc}_N \simeq N^{\theta} v
\label{exciN}
\end{eqnarray}
(For short-ranged models in dimension $d$ where $N=L^d$, the correspondence reads
$\theta=\theta_l/d$)

\subsection{ Difference between different Boundary Conditions in a given sample }

The standard procedure to measure the droplet exponent
is to compute, in each given sample ${\cal J}$, the difference between the ground-state
energies
corresponding to different boundary conditions \cite{mcmillan,bray_moore,fisher_huse}, for
instance Periodic-Antiperiodic
\begin{eqnarray}
  E^{GS(P)}_{\cal J}-E^{GS(AP)}_{\cal J}  = N^{\theta} u
\label{edw}
\end{eqnarray}
where $u$ is an $O(1)$ random variable of zero mean (with a probability distribution
symmetric in $u \to -u$)
Here the droplet exponent has thus the meaning of a Domain-Wall exponent, or stiffness
exponent.
The link with Eq. \ref{jr} is that the energy difference between different boundary conditions
somewhat measures the renormalized coupling between the boundaries.
The link with Eq. \ref{exciN} is that the change of boundary condition will select an
optimized system-size excitation given the new constraints.

For the short-ranged model on hypercubic lattices of dimension $d$, 
the values measured for the stiffness exponent
$\theta_l$ (see \cite{boettcher} and references therein)
reads for the exponent $\theta=\theta_l/d$ defined with respect to the number $N=L^d$ of spins
\begin{eqnarray}
\theta(d=2) && \simeq - \frac{0.28}{2} \simeq -0.14
\nonumber \\
\theta(d=3) && \simeq \frac{0.24}{3} \simeq 0.08
\nonumber \\
\theta(d=4) && \simeq  \frac{0.61}{4} \simeq 0.15
\nonumber \\
\theta(d=5) && \simeq  \frac{0.88}{5} \simeq 0.176
\nonumber \\
\theta(d=6) && \simeq \frac{1.1}{6} \simeq 0.183
\label{hypercubic}
\end{eqnarray}

\subsection{ Role of the droplet exponent in the statistics of the ground state energy  }

The statistics over samples of the ground state energy in spin-glasses has been much studied
recently
(see
\cite{andreanov,Bou_Krz_Mar,pala_gs,aspelmeier_MY,Katz_gs,Katz_guiding,aspelmeier_BMM,boettcher_gs,us_tails,us_matching}
and references therein)
with the following conclusions

(i) the averaged value over samples of the ground state energy reads
\begin{eqnarray}
\overline{ E_{\cal J}^{GS}(N) }  \simeq N e_0+ N^{\theta_{shift}} e_1+...
\label{e0av}
\end{eqnarray}
The first term $N e_0$ is the extensive contribution,
whereas the second term $N^{\theta_{shift}} e_1$ represents the leading correction to
extensivity.

(ii) the fluctuations around this averaged value are governed by some fluctuation exponent
$\mu$
\begin{eqnarray}
E_{\cal J}^{GS}(N)- \overline{ E_{\cal J}^{GS}(N) }  \simeq N^{\mu} u +...
\label{e0fluct}
\end{eqnarray}
where $u$ is an $O(1)$ random variable of zero mean $\overline{u}=0$ by definition.

For spin-glasses in finite dimension $d$, it has been proven that $\mu=1/2$ and that the
distribution of $u$ is simply Gaussian
\cite{aizenman_wehr} suggesting some central Limit theorem coming from the random
couplings.
But the shift-exponent of Eq. \ref{e0av} is non-trivial and coincides with the droplet
exponent \cite{Bou_Krz_Mar}
\begin{eqnarray}
\theta_{shift}= \theta
\label{thetashift}
\end{eqnarray}
The link with Eq \ref{exciN} is that the boundary conditions always induce some
system-size frustration, and thus some system-size excitations.
This contribution of order $N^{\theta}$ is distributed, but the corresponding fluctuations
are sub-leading with respect to the bigger fluctuations corresponding to $\mu=1/2$.

Besides short-ranged models, the fluctuation exponent $\mu$
and the scaling distribution of $u$ have been also studied
for long-ranged models \cite{KY,KKLH,KKLJH}, and the fully connected SK model
\cite{andreanov,Bou_Krz_Mar,pala_gs,aspelmeier_MY,Katz_gs,Katz_guiding,aspelmeier_BMM,boettcher_gs,us_tails,us_matching}.

\section { Brief reminder on the overlap within the replica theory }

\label{app_rep}

Within the replica theory \cite{replica}, the probability distribution  of the cumulative
overlap distribution
\begin{eqnarray}
Y_{\cal J}(q_0) && \equiv \int_{ \vert q \vert \geq q_0} dq P_{\cal J}(q) 
\label{cumulreplicazero}
\end{eqnarray}
has a non-trivial limit in the thermodynamic limit $N=+\infty$ \cite{mezard84,mezard85}:
the translation for the cumulative overlap distribution over the central region
\begin{eqnarray}
I_{\cal J}(q_0) && \equiv \int_{-q_0}^{q_0} dq P_{\cal J}(q) = 1- Y_{\cal J}(q_0)
\label{cumulreplica}
\end{eqnarray}
yields that the probability distribution $\Pi_{\mu}(I)$
is indexed by the parameter 
\begin{eqnarray}
\mu= \mu(\beta,q_0) = 1- \overline{Y_{\cal J}(q_0)} =  \overline{I_{\cal J}(q_0) }
\label{repmu}
\end{eqnarray}
Near the origin $I \to 0^+$, there is the power-law divergence 
\begin{eqnarray}
\Pi_{\mu}(I) \oppropto_{ I \to 0 } I^{- (1-\mu) }
\label{repzero}
\end{eqnarray}
whereas near the other boundary $I \to 1^-$, there is an essential singularity
\begin{eqnarray}
\Pi_{\mu}(I) \oppropto_{ I \to 1} e^{ - \frac{1}{ 4 z_0(\mu) (1-I)} }
\label{repun}
\end{eqnarray}
where $z_0(\mu)$ is given in \cite{mezard84,mezard85}.
Other singularities appear at $(1-I)=1/n$ where $n$ is an integer \cite{Der_Flyv}.

From the point of view of the typical value $I^{typ}$ discussed in the text,
the important point is that it remains finite for $N=+\infty$ 
\begin{eqnarray}
\ln I^{typ} \equiv \overline{\ln I} = \int_0^1 dI (\ln I) \Pi_{\mu}(I)
\label{repityp}
\end{eqnarray}

\end{document}